\documentclass[a4paper,11pt]{article}
\pdfoutput=1 % if your are submitting a pdflatex (i.e. if you have images in pdf, png or jpg format)
\usepackage{jcappub} % for details on the use of the package, please see the JHEP-author-manual
\usepackage[T1]{fontenc} % if needed
\usepackage{natbib}
\usepackage{tensor}
\usepackage{soul,xcolor}
\usepackage{diagbox}
\usepackage{mathrsfs}
\usepackage{tikz}
\usepackage{simplewick}
\usepackage{braket}
\usepackage{comment}

\usetikzlibrary{matrix,shapes,fit,tikzmark,calc,decorations.pathreplacing,decorations.markings,snakes}

\def\cs{|c_s|}
\def\mpl{M_{\rm Pl}}
\usetikzlibrary{decorations.markings,3d,arrows.meta,patterns}
\definecolor{LouisBlue}{RGB}{55, 114, 202}
\definecolor{LouisOrange}{RGB}{180, 54, 22}
\definecolor{LouisColor1}{RGB}{0, 118, 63}
\definecolor{LouisColor2}{RGB}{111, 73, 189}
\def\propagator{
\begin{tikzpicture}[thick]
\clip (-1.51,-0.4)rectangle(1.51,0.4);
\draw[{Square[open]}-{Square[open]}](-1.5,0)--(1.5,0);
\draw[] (-1.,-0.3)node{\small$\pm$};
\draw[] (1.,-0.3)node{\small$\pm$};
\end{tikzpicture}
}

\def\ba{\begin{eqnarray}}
\def\ea{\end{eqnarray}}
\newcommand{\D}{{\rm d}}

\newcommand{\beq}{\begin{equation}}
\newcommand{\eeq}{\end{equation}}
\newcommand{\bal}{\begin{aligned}}
\newcommand{\eal}{\end{aligned}}

\renewcommand{\vec}[1]{\mathbf{#1}}

\title{\boldmath Loops in inflation with strongly non-geodesic motion}

\author[a]{Sebastian Garcia-Saenz,}
\author[a,b,c]{Yizhou Lu,}
\author[d]{S\'ebastien Renaux-Petel}

\affiliation[a]{
Department of Physics, Southern University of Science and Technology, \\
Shenzhen 518055, China
}

\affiliation[b]{
Shanghai Institute for Mathematics and Interdisciplinary Sciences, Shanghai 200000, China
}
\affiliation[c]{
School of Mathematical Sciences, Fudan University, Shanghai 200000, China
}
\affiliation[d]{
Institut d'Astrophysique de Paris, UMR 7095 du CNRS et de Sorbonne Universit\'e, 98 bis Boulevard
Arago, 75014 Paris, France
}

\emailAdd{sgarciasaenz@sustech.edu.cn}
\emailAdd{luyz@simis.cn}
\emailAdd{renaux@iap.fr}

\abstract{
We study loop corrections in the effective field theory of inflation with imaginary speed of sound, which has been shown to provide an effective description of multi-field inflationary models characterized by strongly non-geodesic motion and heavy entropic perturbations. We focus on the one-loop corrections to the scalar and tensor power spectra, taking into account all relevant vertices at leading order in derivatives and in slow-roll. We find a power-law dependence of the scalar two-point function on the scale that defines the range of validity of the effective theory, analogous to the enhancement observed in tree-level correlation functions. Even more dramatic, the relative correction to the tensor spectrum is exponentially enhanced, albeit also suppressed in the slow-roll limit. In spite of these large effects, our results show that this class of models can satisfy the requirement of perturbative control and a consistent loop expansion within a range of parameters of phenomenological interest. On the other hand, models predicting large values of the power spectrum on small scales are found to be under strong tension. As a technical bonus, we carefully explain the prescription for the regularization and manipulation of loop integrals in this set-up, where one has a non-trivial domain of integration for time and momentum integrals owing to the regime of validity of the effective field theory. This procedure is general enough to be of potential applicability in other contexts.
}

\begin{document} 
\maketitle
\flushbottom

\section{Introduction}

Scalar and tensor inhomogeneities on small scales --- i.e.\ wavelengths shorter than those probed in the cosmic microwave background (CMB) and large-scale structure surveys --- are a primary target of current and future cosmological experiments. A positive detection could potentially provide invaluable information about the physics of the dark era of inflation following the horizon crossing of CMB-scale fluctuations, i.e.\ a large number of e-folds of evolution about which we have essentially no experimental knowledge so far.

There is no good reason to expect the near-perfect scale invariance of perturbations observed in the CMB to extrapolate until the end of inflation. Transient violations of slow-roll, triggered for instance by features in the scalar potential, see e.g.~\cite{Kohri:2007qn,Kawasaki:2016pql,Garcia-Bellido:2017mdw,Kannike:2017bxn,Germani:2017bcs,Motohashi:2017kbs,Di:2017ndc,Ballesteros:2017fsr,Hertzberg:2017dkh,Ozsoy:2018flq,Ballesteros:2018wlw,Cai:2019bmk,Lin:2020goi,Ballesteros:2020qam,Palma:2020ejf,Fumagalli:2020adf,Fumagalli:2020nvq,Braglia:2020taf,Fumagalli:2021cel,Dalianis:2021iig,Fumagalli:2021mpc,Pi:2022ysn,Karam:2022nym,Inomata:2022yte,Fumagalli:2023loc,Caravano:2024tlp,Caravano:2024moy,Briaud:2025hra,Caravano:2025diq}, may enhance scalar inhomogeneities to observable levels, for example through their effect on primordial black hole formation \cite{Hawking:1971ei,Carr:1974nx,Carr:1975qj}. 
Moreover, scalar fluctuations also source tensor modes, i.e.\ induced gravitational waves (GWs) \cite{Mollerach:2003nq,Ananda:2006af,Baumann:2007zm,Kohri:2018awv,Espinosa:2018eve,Cai:2018dig}, which could be directly detected by next-generation experiments and might have in fact already been discovered in pulsar timing array data \cite{Vagnozzi:2023lwo,Cai:2023dls,Inomata:2023zup,Wang:2023ost,Liu:2023ymk,Ebadi:2023xhq,Figueroa:2023zhu,Yi:2023mbm,Liu:2023pau,Domenech:2024rks}. Even on CMB scales, actually, departures from the `vanilla'-type scenario of single-field slow-roll inflation are of course also interesting even if already constrained by observations. This is particularly relevant in relation to the higher-point statistics of the perturbations, i.e.\ the non-Gaussianities, which encode information on the microscopic physics of inflation and its particle content \cite{Chen:2010xka,Wands:2010af,Wang:2013zva,Renaux-Petel:2015bja,Achucarro:2022qrl}. There is therefore a strong motivation to thoroughly understand, on the theoretical side, the physical mechanisms that could give rise to a magnification of scalar perturbations and to large non-Gaussianities. 
Multi-field inflation provides a natural arena to investigate this question in view of the wealth of possible interesting dynamics that would not be available in the case of a single degree of freedom. Two uniquely multi-field features that are worth highlighting are the (generic) presence of a curved field space, i.e.\ a non-linear sigma model where the kinetic term of the scalar fields $\phi^I$ has the form $G_{IJ}(\phi)\nabla^{\mu}\phi^I\nabla_{\mu}\phi^J$ with non-flat `internal' metric $G_{IJ}$, and the possibility that the inflaton trajectory may deviate from a geodesic in the space spanned by the fields $\phi^I$.

In fact, in negatively curved field spaces, slow-roll trajectories tend to be geometrically destabilized, in which case the inflationary fields bifurcate into another attractor regime characterized by a strongly non-geodesic motion \cite{Renaux-Petel:2015mga}. This mechanism to inflate is radically different from slow-roll, it naturally predicts large non-Gaussianities and it has been extensively studied in the past few years, see e.g.~\cite{Cremonini:2010ua,Renaux-Petel:2015mga,Brown:2017osf,Renaux-Petel:2017dia,Mizuno:2017idt,Achucarro:2017ing,Christodoulidis:2018qdw,Linde:2018hmx,Garcia-Saenz:2018ifx,Garcia-Saenz:2018vqf,Achucarro:2018vey,Christodoulidis:2019mkj,Bjorkmo:2019aev,Garcia-Saenz:2019njm,Fumagalli:2019noh,Bjorkmo:2019fls,Christodoulidis:2019jsx,Bjorkmo:2019qno,Aragam:2019omo,Bravo:2020wdr,Chakraborty:2019dfh,Ferreira:2020qkf,Aragam:2020uqi,Aragam:2021scu,Renaux-Petel:2021yxh,Anguelova:2022foz,Christodoulidis:2023eiw,Anguelova:2024akm,Wolters:2024vzk}. Considering for concreteness an adiabatic-entropic decomposition of the scalar fields' perturbations \cite{Gordon:2000hv} and assuming the well-motivated scenario in which the entropic modes are heavy (in a sense to be made more precise below), one is led to an effective single-field description
in which the adiabatic mode propagates with sub-luminal speed of sound given by \cite{Tolley:2009fg,Achucarro:2010da}
\begin{equation}
\frac{1}{c_s^2}=1+\frac{4H^2\eta_\perp^2}{m_s^2} \,,
\end{equation}
where $H$ is the Hubble scale, $\eta_\perp$ is a dimensionless parameter that quantifies the degree of geodesic deviation and $m_s^2$ is a scale related to the effective mass of the entropic perturbations (see e.g.\ Ref.\ \cite{Garcia-Saenz:2019njm} for the precise definitions and further details). Note that all these quantities are in general time-dependent, although in this paper we will assume throughout adiabaticity conditions \cite{Cespedes:2012hu,Achucarro:2012yr} and thus approximate them as constants. By definition, we speak of a strongly non-geodesic field trajectory as one with large bending parameter, $|\eta_\perp|\gg1$. More precisely, for reasons we explain below, in this paper we will be interested in a regime where $\eta_\perp^2\gtrsim m_s^2/H^2\gg 1$. This condition implies that $c_s^2$, when positive, must be parametrically smaller than unity. As is well known, this scenario of reduced speed of sound leads to large non-Gaussianities as one of its key predictions \cite{Chen:2006nt,Cheung:2007st}.

In slow-roll and slow-turn inflation, i.e.\ with $|\eta_\perp|\lesssim 1$, the entropic mass squared, $m_s^2$, is always a positive quantity. 
However, $m_s^2$ is a derived parameter, unrelated to the mass of any fundamental particle, and in general it can have either sign. 
Explicitly, in non-linear sigma models with two fields, one has
\begin{equation}
m_s^2=V_{;ss}-H^2\eta_\perp^2+\epsilon H^2M_{\rm Pl}^2R_{\rm fs} \,,
\end{equation}
where $V_{;ss}$ is the second field space-covariant derivative of the potential projected along the entropic direction, $M_{\rm Pl}$ is the Planck mass and $R_{\rm fs}$ is the curvature scalar of the internal space. 
The latter of course need not be positive, and in fact negatively curved field spaces are especially well-motivated by constructions of inflationary models within supergravity \cite{Kallosh:2013yoa,Carrasco:2015uma,Renaux-Petel:2015mga,Achucarro:2017ing}. 
More to the point, the bending parameter always gives a negative contribution to $m_s^2$, and indeed several explicit models are characterized by background solutions with $m_s^2<0$, see e.g.\ \cite{Renaux-Petel:2015mga,Christodoulidis:2018qdw,Garcia-Saenz:2018ifx,Fumagalli:2019noh}.
Beyond two fields, the entropic mass (in general a matrix) also includes terms related to the `twist' of the adiabatic-entropic basis, and these too will typically give negative contributions \cite{Pinol:2020kvw,Aragam:2019omo,Aragam:2020uqi,Christodoulidis:2022vww,Christodoulidis:2023eiw}.

Even though $m_s^2$ does not properly correspond to any mass eigenstate, the intuition that a negative squared mass is associated to a tachyonic instability may be shown to be correct. Explicit computations indeed indicate, in this class of set-ups, that perturbations experience an exponential enhancement starting around the time of `mass-shell' crossing, $k^2/a^2\sim |m_s^2|$ \cite{Garcia-Saenz:2018ifx,Garcia-Saenz:2018vqf}. 
At earlier times, i.e.\ for larger physical momenta, the instability is absent, as expected of tachyonic modes. Unlike for tachyonic fields in flat spacetime, however, the instability is also quenched soon after horizon crossing. This is so because, in the super-Hubble regime, entropic fluctuations evolve independently of adiabatic ones, and a good notion of effective mass may be defined in this context. This mass is however different from $m_s^2$; specifically in two-field models it is given by $m^2_{s,{\rm eff}}=m_s^2+4H^2\eta_\perp^2$. We see therefore that a large bending has the ability to both trigger a transient tachyonic destabilization while at the same time to render the super-horizon fluctuations (and hence the background in particular) stable. In fact, since typically one has $m^2_{s,{\rm eff}}\gg H^2$ in scenarios with strongly non-geodesic motion, entropic modes quickly decay after horizon crossing and the curvature perturbation reaches an adiabatic limit.

A tachyonic instability during inflation leads to several interesting observational signatures. At the level of the two-point function, the exponential enhancement of scalar fluctuations translates into a very small tensor-to-scalar ratio, so that a detection of tensor modes in the CMB could easily falsify this scenario. 
More promising and intriguing however is the possibility of probing inflationary tachyonic modes through higher-point statistics. 
Non-Gaussianities in this context have been shown to be large, as in reduced speed of sound scenarios, but with distinctive shapes: the bispectrum overlaps strongly with the orthogonal template \cite{Garcia-Saenz:2018vqf}, while higher-point correlators are similarly enhanced on `flattened' polygon configurations \cite{Fumagalli:2019noh,Bjorkmo:2019qno}.

Although a tachyonic instability in inflation is necessarily transient, as we have explained, the exponential growth still carries the risk of impairing the perturbative description. This question motivates us to tackle the problem of assessing the size of loop corrections in models with tachyonic modes. Concretely, we want to see under what conditions the one-loop corrections to the scalar power spectrum is smaller than the tree-level result. Here `loops' are understood in the sense of the in-in formalism \cite{Weinberg:2005vy} and hence include both classical and quantum effects.

We are also interested in assessing the size of scalar loop corrections to the tensor power spectrum. In this case, a large one-loop contribution need not be associated with the breakdown of perturbativity, since one may still a priori have a consistent expansion for the higher-order loops. This could result in a significant contribution to scalar-induced GWs during inflation
\cite{Fumagalli:2021mpc,Inomata:2021zel,Ota:2022hvh,Ota:2022xni,Feng:2023veu}. In this event, a tachyonic instability could serve as a mechanism to enhance tensor fluctuations to levels of interest for current and future GW detectors. Beyond this interesting phenomenological aspect, and more to the point of this paper, the tensor spectrum is also another observable that may be used for diagnosing the perturbativity of the theory.

Having established these goals, one is faced with the technically challenging task of calculating loop integrals in complex multi-field models in which the mode functions are not known analytically. In this paper we bypass this issue by resorting to the effective field theory (EFT) of single-field inflation \cite{Cheung:2007st}, assuming further the limits of slow-roll and adiabaticity. Besides the fact that this approach will allow us to obtain fully analytical expressions (with some approximations and assumptions to be made explicit below), it has the added advantage that it provides results that are universal, within its regime of validity, as is usual with EFT.

The EFT of inflation derived from integrating out a tachyonic ($m_s^2<0$) and heavy ($|m_s^2|\gg H^2$) mode was introduced in \cite{Garcia-Saenz:2018ifx} and further studied in \cite{Garcia-Saenz:2018vqf,Fumagalli:2019noh,Ferreira:2020qkf,Ballesteros:2021fsp} (see also \cite{Cremonini:2010ua} for an earlier work). The structure of the theory is equivalent to that of the standard EFT of \cite{Cheung:2007st}, the only peculiarity being that the squared speed of sound, $c_s^2$, of the adiabatic fluctuation is negative in this set-up, hence the so-called scenario of imaginary speed of sound. If $c_s$ is imaginary, the mode function is not oscillatory but exponentially growing or decaying, which is nothing but a manifestation of the tachyonic instability of the multi-field theory. Although the instability in the EFT may seem more drastic (a wrong-sign gradient term versus a wrong-sign mass term in the case of a tachyon), one must remember that any EFT is self-consistently endowed with an energy cutoff, which in the present context naturally limits the regime of validity of the theory to capture the physics around the time of sound horizon crossing. In this regime, then, a breakdown of perturbativity in the multi-field theory should imply an equivalent breakdown in the single-field EFT. Remarkably, at \textit{tree level}, $n$-point correlation functions of the adiabatic fluctuation have been shown to admit a consistent perturbative expansion \cite{Fumagalli:2019noh}. As explained, one of our goals here is to address this question at the one-loop level.

Sec.\ \ref{sec:preliminary} contains all the background material we will make use of in our calculations, including a brief review of the EFT of inflation with imaginary sound speed, of the in-in formalism, and a discussion on the manipulation of the loop integrals in the presence of an explicit momentum cutoff. To our knowledge this last aspect has not been addressed previously in the literature, which we think may prove useful in other contexts. Sec.\ \ref{sec:one loop scalar spectrum} presents our calculation of the one-loop correction to the scalar power spectrum. Our results are fully analytic although not exhaustive, as we consider the approximation of large $x$ (the dimensionless parameter defining the momentum cutoff of the EFT, to be defined explicitly below) and moreover we do not calculate all the possible diagrams. Nevertheless, we also provide a scaling argument which shows that the neglected diagrams must yield similar results and would therefore not modify our final estimates. The same calculation for the tensor power spectrum is given in Sec.\ \ref{sec:one loop tensor spectrum}, using the same large-$x$ approximation.
The conclusions that follow from our results, together with an outlook of our paper, are discussed in Sec.\ \ref{sec:discussion}. Some technical aspects and details of our calculations are relegated to the appendices.

%%%%%%%%%%%%%%%%%%%%%%%%%%%%%%%%
%%%%%%%%%%%%%%%%%%%%%%%%%%%%%%%%

\section{EFT with imaginary speed of sound and loop integrals} \label{sec:preliminary}

\subsection{Quadratic action and quantization}
\label{subsec:quantization}

The construction of the EFT of inflation with imaginary speed of sound is equivalent to that of the standard set-up \cite{Cheung:2007st} (see also \cite{Piazza:2013coa} for a review). Throughout this paper we focus exclusively on the EFT at lowest order in the derivative expansion and leading order in the slow-roll approximation. This set-up is thus equivalent to the theory of k-inflation \cite{Garriga:1999vw,Armendariz-Picon:2000nqq} at lowest order in the slow-roll expansion; see Appendix \ref{app:k-inf} for details.

At quadratic order in perturbations one has the Lagrangian
\begin{equation} \label{eq:quad action}
    S_{2}=\int\D\eta\D^3\vec x\, a^2\left[ \frac{\epsilon M_{\rm Pl}^2}{c_s^2}\left(
    \zeta'^2-c_s^2(\partial\zeta)^2
    \right)+\frac{M_{\rm Pl}^2}{8}
    \left(\gamma'_{ij}\gamma'_{ij}-\partial_k\gamma_{ij}\partial_k\gamma_{ij}\right)
    \right] \,,
\end{equation}
for the adiabatic curvature perturbation $\zeta$ and the transverse and traceless metric perturbation $\gamma_{ij}$. Here $a(\eta)$ is the scale factor defined in terms of conformal time $\eta$, and primes denote derivatives with respect to $\eta$. $M_{\rm Pl}$, $\epsilon$ and $c_s^2$ are respectively the Planck mass, first slow-roll parameter and squared speed of sound. Both $\epsilon$ and $c_s^2$ are assumed constant at the order we work on in the slow-roll approximation,\footnote{In the multi-field picture, this means that our description applies to situations with strong but approximately constant turn, as it occurs in rapid-turn attractor models, e.g.\ \cite{Renaux-Petel:2015mga,Brown:2017osf,Garcia-Saenz:2018ifx}. Scenarios with strong and sharp turns are therefore not captured by our approximation, where even a single-field EFT treatment may be inapplicable.} and consistently with this we have $a(\eta)=-1/(H\eta)$, with $H$ the Hubble parameter.

As explained in the Introduction, $c_s^2$ is a derived quantity related to parameters of the multi-field theory, and may actually be negative. This is the scenario we consider in this paper, and from now on we write $c_s^2=-|c_s|^2$ to be explicit. From \eqref{eq:quad action} we infer that $\zeta$ is then a ghost field (wrong-sign kinetic term) and gradient-unstable (exponentially growing solutions). Nevertheless, keeping in mind that this is an EFT with a limited regime of validity (to be discussed below), nothing prevents one from carrying out the canonical quantization as usual. This was done in \cite{Garcia-Saenz:2018vqf} and we review this procedure next.

Promoting $\zeta$ to a quantum field, it is decomposed in Fourier modes as
\begin{equation}
    \hat{\zeta}(\eta,\vec x)=\int\frac{\D^3\vec k}{(2\pi)^3}
    \hat{\zeta}_{\vec k}(\eta)e^{i\vec k\cdot \vec x} \,,
\end{equation}
with
\begin{equation}
\hat{\zeta}_{\vec k}=\zeta_{k
    }(\eta)a_{\vec k}+\zeta_{k}^*(\eta)a_{-\vec k}^\dagger \,,
\end{equation}
in terms of the mode function $\zeta_k(\eta)$ and the annihilation and creation operators $\hat a$ and $\hat a^\dagger$. The latter satisfy the commutation relations
\begin{equation}
    \left[
    \hat{a}_{\vec k},\hat{a}_{\vec p}^\dagger
    \right]=(2\pi)^3\delta ^3(\vec k-\vec p) \,,\qquad 
    \left[
    \hat{a}_{\vec k},\hat{a}_{\vec p}
    \right]=\left[
    \hat{a}_{\vec k}^\dagger,\hat{a}_{\vec p}^\dagger
    \right
    ]=0 \,.
\end{equation}
The free-theory equation of motion for $\zeta_k$ that follows from \eqref{eq:quad action} reads
\begin{equation}
    \zeta_k''-\frac{2}{\eta}\zeta_k'-\cs^2 k^2\zeta_k=0 \,,
\end{equation}
with general solution
\begin{equation}
\zeta_k(\eta)=\zeta_{k,+}(\eta)+\zeta_{k,-}(\eta) \,,
\end{equation}
and we introduce what we call the growing and decaying modes, respectively given by
\begin{equation} \label{eq:growing-decaying modes}
\zeta_{k,+}\equiv \frac{A_k}{k^{3/2}}e^{k|c_s|\eta}(k|c_s|\eta-1) \,,\qquad \zeta_{k,-}\equiv -\frac{B_k}{k^{3/2}}e^{-k|c_s|\eta}(k|c_s|\eta+1) \,.
\end{equation}
Here $A_k$ and $B_k$ are integration constants, a priori dependent on $k$; the reason for extracting the factors of $k^{3/2}$ will become clear in a moment.

Let $\mathscr P_\zeta(\eta,\vec x)\equiv \partial\mathcal L_2/\partial\zeta'$ be the canonical momentum. The quantization condition $[\hat\zeta(\eta,\vec x),\hat{\mathscr P}_{\zeta}(\eta,\vec y)]=i\delta^3(\vec x-\vec y)$ then translates into
\begin{equation}
    \zeta_k\zeta_k^{*\prime}-\zeta_k'\zeta_k^*=-\frac{i\cs^2}{2\epsilon a^2M_{\rm Pl}^2} \,,
\end{equation}
or equivalently
\begin{equation} \label{eq:wronskian AB}
    \mathrm{Im}[A_k^*B_k]=\frac{H^2}{8\cs M_{\rm Pl}^2\epsilon} \,.
\end{equation}
This last relation shows that both the growing and decaying modes must be present, unlike what occurs in the standard set-up with real $c_s$, where one typically chooses one solution corresponding to the Bunch-Davies vacuum.

Without loss of generality we choose $A_k$ real, which means that $\mathrm{Im}[B_k]\neq 0$, and we write
\begin{equation}
    A_k=\alpha_ke^x \,,\qquad B_k=\alpha_k \rho_k e^{i\psi_k}e^{-x} \,,
\end{equation}
where all parameters, $\alpha_k$, $\rho_k$ and $\psi_k$, are real. These are not independent but are constrained by Eq.\ \eqref{eq:wronskian AB}, explicitly
\begin{equation}\label{eq:wronskian}
    \alpha_k^2\rho_k\sin\psi_k=\frac{H^2}{8\cs M_{\rm Pl}^2\epsilon} \,.
\end{equation}
As the right-hand side of this equation is scale invariant at the order in the slow-roll approximation that we assume, one has generically that $\alpha_k$, $\rho_k$ and $\psi_k$ must likewise be scale invariant. So from now on we write $\alpha_k=\alpha$, $\rho_k=\rho$ and $\psi_k=\psi$, as well as $A_k=A$ and $B_k=B$ (we will continue to use the latter in some intermediate results).

We have also introduced the parameter $x$, which is actually redundant as it may be absorbed into the definitions of $\alpha$ and $\rho$, but is nonetheless useful as a measure of the amplification of the power spectrum resulting from the tachyonic instability (see Eq.\ \eqref{eq:tree scalar spectrum} below). From \eqref{eq:growing-decaying modes} we then infer that both modes had comparable size (assuming $\rho$ of order unity) at the time $\eta$ such that $-k\cs\eta=x$ (for any given $k$). We interpret this as defining the time at which the EFT starts to be valid, so that $x$ is essentially a dimensionless measure of the UV cutoff of the theory.

In the diagrammatic organization of our calculations it will be useful to treat $\zeta_{k,+}$ and $\zeta_{k,-}$ separately. In this vein we define four distinct propagators:
\begin{equation}
\braket{\zeta_{\pm,\mathbf k}(\eta)\zeta_{\pm,\mathbf p}(\eta)}\quad=\quad\parbox{100pt}{\propagator} \equiv \quad \zeta_{k,\pm}(\eta)\zeta_{p,\pm}^*(\eta)(2\pi)^3\delta^3(\vec k+\vec p) \,.
\end{equation}
The tree-level, dimensionless scalar power spectrum is given by\footnote{We use the usual notation where a prime on a correlator means that a factor of $(2\pi)^3$ and momentum-conserving delta function have been removed.}
\begin{equation} \label{eq:tree scalar spectrum}
\mathcal P_{\zeta}^{(\rm tree)}\equiv \frac{k^3}{2\pi^2}\braket{\hat\zeta_{\vec k}\hat\zeta_{\vec p}}'\Big|_{\eta\to 0^-}=\frac{k^3}{2\pi^2}|\zeta_k|^2\simeq\frac{\alpha^2}{2\pi^2}e^{2x} \,,
\end{equation}
where we have taken the late-time limit and the last approximation is valid in the regime of large $x$.

We next consider the quantization of the graviton field. This case is of course completely standard but we include it anyway for completeness. The quantum operator $\hat\gamma_{ij}$ is expanded in Fourier modes as
\begin{equation}
    \hat\gamma_{ij}(\eta,\vec x)=\int\frac{\mathrm d^3\mathbf k}{(2\pi)^3}e^{i\mathbf k\cdot \mathbf x}\hat\gamma_{ij,\mathbf k}(\eta) \,,
\end{equation}
where
\begin{equation}
    \hat{\gamma}_{ij,\mathbf{k}}(\eta)=\sum_{\lambda=+,-}\varepsilon_{ij}^\lambda(\hat{\mathbf  k})\gamma_k(\eta)\hat{a}_{\mathbf{k}}^\lambda+\varepsilon_{ij}^{\lambda*}(-\hat{\mathbf k})\gamma_k^*(\eta)\hat{a}_{-\mathbf{k}}^{\lambda\dagger} \,,
\end{equation}
and $\hat{\vec k}\equiv \vec k/k$. The index $\lambda$ denotes the polarization of the mode, and the polarization tensors $\varepsilon^\lambda_{ij}$ satisfy
\begin{equation}
    k_i\varepsilon_{ij}^\lambda(\hat{\vec k})=0=\varepsilon_{ii}^\lambda \,,\qquad \varepsilon_{ij}^{\lambda*}(\hat{\vec k})\varepsilon_{ij}^{\sigma}(\hat{\vec k})=\delta^{\lambda\sigma} \,,
    \qquad \varepsilon_{ij}^\lambda(\hat{\vec k})=\epsilon_{ij}^{\lambda*}(-\hat{\vec k}) \,.
\end{equation}
The canonical commutation relations obeyed by $\hat{a}_{\vec k}^\lambda$ and $\hat{a}_{\vec k}^{\lambda\dagger}$ are
\begin{equation}
    \left[\hat{a}_{\mathbf k}^\lambda,\hat{a}_{\mathbf p}^{\sigma\dagger}\right]=(2\pi)^3\delta^{\lambda\sigma}\delta^3(\mathbf k-\mathbf p) \,,\qquad 
    \left[
    \hat{a}_{\vec k}^{\lambda},\hat{a}_{\vec p}^{\sigma}
    \right
    ]=
    \left[
    \hat{a}_{\vec k}^{\lambda\dagger},\hat{a}_{\vec p}^{\sigma\dagger}
    \right]=0 \,.
\end{equation}
The quantization condition $[\hat\gamma_{ij}(\eta,\vec x),\hat{\mathscr P}_{\gamma ij}(\eta,\vec y)]=i\delta^3(\vec x-\vec y)$ gives
\begin{equation}\label{eq:wronskian_tensor}
    \gamma_k\gamma_k^{\prime*}-\gamma_k'\gamma_k^*=\frac{4i}{a^2M_{\rm Pl}^2} \,,
\end{equation}
and the mode functions satisfy the free-theory equation of motion
\begin{equation}
    \gamma_k''-\frac{2}{\eta}\gamma_k'+k^2\gamma_k=0 \,,
\end{equation}
working in the de Sitter limit. The general solution is
\begin{equation}
    \gamma_k(\eta)=\tilde A_k(-k\eta)^{3/2}H_{3/2}^{(1)}(-k\eta)+\tilde B_k(-k\eta)^{3/2}H_{3/2}^{(2)}(-k\eta) \,,
\end{equation}
in terms of the Hankel functions $H^{(1)}_\nu$ and $H^{(2)}_\nu$. The choice of Bunch-Davies vacuum fixes
\begin{equation}
    \tilde A_k=\frac{\sqrt\pi H}{k^{3/2}M_{\rm Pl}},\quad \tilde B_k=0 \,,
\end{equation}
and therefore
\begin{equation} \label{eq:tensorBDmode}
    \gamma_k(\eta)=\frac{\sqrt{\pi}H}{k^{3/2}M_{\rm Pl}}(-k\eta)^{3/2}H_{3/2}^{(1)}(-k\eta) \,.
\end{equation}

The two-point correlator or propagator is given by
\begin{equation}
    \braket{\gamma_{ij,\mathbf k}(\eta)\gamma_{kl,\mathbf p}(\eta)}\quad=\quad
    \parbox{70pt}{
\begin{tikzpicture}[thick]
\draw[decorate,decoration={snake},double](-1,0)--(1,0);
\draw[fill=white](-1.1,0.1)--(-0.9,0.1)--(-0.9,-0.1)--(-1.1,-0.1)--(-1.1,0.1);
\draw[fill=white](0.9,-0.1)rectangle(1.1,0.1);
\end{tikzpicture}
    }\quad =\quad \textbf{}
    \gamma_k(\eta)\gamma_k^*(\eta) P_{ij,kl}(\hat{\mathbf k})(2\pi)^3\delta^3(\mathbf k+\mathbf p) \,,
\end{equation}
where the projection tensor is defined by
\begin{equation}
     P_{ij,kl}(\hat{\mathbf k})\equiv \sum_\lambda\varepsilon_{ij}^\lambda(\hat{\mathbf k})\varepsilon_{kl}^{\lambda*}(\hat{\mathbf k})=\frac12( P_{ik} P_{jl}+ P_{il} P_{jk}- P_{ij} P_{kl}) \,,\qquad  P_{ij}\equiv \delta_{ij}-\hat{k}_i\hat{k}_j \,.
\end{equation}
Finally, the tree-level tensor power spectrum is given by
\begin{equation}
    \mathcal P_\gamma^{(\rm tree)}=\frac{k^3}{\pi^2}|\gamma_k|^2\Big|_{\eta\to0^-}=\frac{2}{\pi^2}\left(
\frac{H}{M_{\rm Pl}}
\right)^2 \,,
\end{equation}
where we have summed over the two polarizations and taken the late-time limit.\footnote{The sum over polarizations is equivalent to tracing over the indices in the correlator, i.e.\ $\langle\gamma_{ij}\gamma_{ij}\rangle'=2|\gamma_k|^2$, since $P_{ij,ij}(\hat{\vec k})=2$. More generally, i.e.\ beyond tree-level, the total dimensionless tensor spectrum is given by $\mathcal{P}_\gamma=\frac{k^3}{2\pi^2}\langle\gamma_{ij}\gamma_{ij}\rangle'$.}

\subsection{Interaction vertices}

The one-loop scalar and tensor spectra receive contributions from diagrams involving the cubic and quartic couplings of the theory. Recall that we focus on the slow-roll approximation, however as we are interested in couplings mixing the adiabatic mode with gravity, we do not strictly take the so-called decoupling limit, but instead keep the leading-order vertices of each type. At cubic order we consider
\begin{equation} \label{eq:cubic action}
S_3=-\int\D\eta\D^3\vec x\, a \frac{M_{\rm Pl}^2\epsilon}{H|c_s|^2}\left(1+|c_s|^2\right)\left[-\frac{\mathcal{A}}{|c_s|^2}\zeta'^3+\zeta'(\partial\zeta)^2\right]-\int\D \eta \D^3 \vec x\, a^2 \epsilon \mpl^2 \gamma_{ij}\partial_i\zeta\partial_j\zeta \,.
\end{equation}
Here the first terms correspond to the $\zeta\zeta\zeta$ vertices appearing in the EFT of inflation at leading order in the slow-roll approximation, with $\mathcal{A}$ a dimensionless constant. Similarly, the second term is the leading-order $\gamma\zeta\zeta$ vertex, corresponding to the universal minimal gravitational coupling. As already mentioned, this action also coincides with the cubic-order expansion of k-inflation \cite{Chen:2006nt,Chen:2010xka} (see Appendix \ref{app:k-inf} for details). For later use we list here the interaction Hamiltonians for the individual vertices:\footnote{We define the Hamiltonian with respect to conformal time $\eta$.}
\begin{align}
H_{\zeta\zeta\zeta}^{(1)}(\eta)&= \mathscr{C}a\int\mathrm d^3\vec x\,\zeta'^3\,, \qquad \mathscr{C}\equiv -\frac{M_{\rm Pl}^2\epsilon}{H}\left(
1+\frac{1}{|c_s|^2}
\right)\frac{\mathcal{A}}{|c_s|^2} \,, \label{eq:cubicHam} \\
H_{\zeta\zeta\zeta}^{(2)}(\eta)&= \tilde{\mathscr C}a\int \D ^3\vec x\,\zeta'(\partial\zeta)^2 \,, \qquad \tilde{\mathscr{C}}\equiv \frac{M_{\rm Pl}^2\epsilon}{H}
\left(
1+\frac{1}{\cs^2}
\right) \,, \label{eq:cubicHam2}
\\
H_{\gamma\zeta\zeta}(\eta)&={\mathscr E}a^2\int \D^3\vec x\,\gamma_{ij}\partial_i\zeta\partial_j\zeta\, ,\qquad \mathscr E\equiv M_{\rm Pl}^2 \epsilon \,. \label{eq:cubicHamMix}
\end{align}

At quartic order in perturbations, and always at leading order in slow-roll, the EFT of inflation contains three $\zeta\zeta\zeta\zeta$ vertices. For simplicity, and in line with our aim of estimating the size of loop corrections rather than calculating exact values, we focus here on a single vertex, namely $\zeta^{\prime 4}$. We do not expect the other two vertices to give qualitatively different results, and indeed we argue in Appendix \ref{app:other quartic} that the one-loop scalar power spectrum for all three vertices should have the same scaling with $x$ in the large-$x$ limit, which we moreover confirm through an explicit calculation with the $(\partial\zeta)^4$ vertex. In addition, we have three mixed quartic couplings $\gamma\gamma\zeta\zeta$ at leading order in derivatives and slow-roll, leading to the following set of interaction Hamiltonians:
\begin{align}
    H_{\zeta\zeta\zeta\zeta}^{(1)}(\eta)&={\mathscr D}\int\mathrm d^3\vec x \,\zeta'^4 \,,\qquad \mathscr D\equiv\frac{M_{\rm Pl}^2\epsilon}{H^2|c_s|^6}\mathcal D \,, \label{eq:quarticHam}\\
    H_{\gamma\gamma\zeta\zeta}^{(1)}(\eta)&=\mathscr F a^2\int \D^3\vec x\, \gamma_{ij}^2\zeta'^2 \,,\qquad \mathscr F\equiv \frac{\mpl^2\epsilon}{\cs^2} \mathcal F \,, \label{eq:quarticHam tensor1}\\
    H_{\gamma\gamma\zeta\zeta}^{(2)}(\eta)&=\tilde{\mathscr F}a^2\int\D^3 \vec x \,\gamma_{ij}^2 (\partial\zeta)^2 \,,\qquad \Tilde{\mathscr F}\equiv \mpl^2  \epsilon\tilde{\mathcal F} \,,\label{eq:quarticHam tensor2}\\
H_{\gamma\gamma\zeta\zeta}^{(3)}(\eta)&=\overline{\mathscr{F}}a^2 \int \mathrm{d}^3 \vec x\, \gamma_{il}\gamma_{jl}\partial_i \zeta \partial_j \zeta \,,\qquad \overline{\mathscr{F}}\equiv M_{\rm Pl}^2\epsilon\overline{\mathcal F} \,. \label{eq:quarticHam tensor3}
\end{align}
Here $\mathcal D$, $\mathcal F$, $\tilde{\mathcal F}$ and $\overline{\mathcal F}$ are some dimensionless constants, generically of $\mathcal{O}(1)$ in the EFT context.\footnote{More in detail, we expect $\mathcal D$, $\mathcal F$, $\tilde{\mathcal F}$ and $\overline{\mathcal{F}}$ to be combinations of $\cs^2$ and order-one constants, see e.g.~\cite{Renaux-Petel:2013wya} for $\mathcal D$. Barring fine tuning, they are therefore $\mathcal{O}(1)$ if $\cs\lesssim1$, but not if $\cs\gg1$ (see footnote \ref{footnote:large cs} in relation to this point). This minor caveat is anyway unimportant for us, since we will see that the quartic vertices only play a minor role in our main results.} In principle they could be computed explicitly in terms of Wilson coefficients appearing in the EFT Lagrangian, however we have not carried out this cumbersome calculation, since anyway we are not interested in $\mathcal{O}(1)$ numbers. Notice incidentally that the last two terms, $H_{\gamma\gamma\zeta\zeta}^{(2)}$ and $H_{\gamma\gamma\zeta\zeta}^{(3)}$, are universal contributions as they arise from the minimal gravitational coupling.

\subsection{In-in formalism}

We employ the standard in-in formalism \cite{Weinberg:2005vy} in our computations (see e.g.\ \cite{Chen:2010xka,Wang:2013zva} for reviews). The master formula giving the vacuum expectation value of an operator $\hat Q$ is
\begin{equation}
    \braket{\hat Q(\eta)}=\bra{0}
    U_{\rm int}^\dagger(\eta,\eta_0) \hat Q_I(\eta) U_{\rm int}(\eta,\eta_0)\ket{0} \,,
\end{equation}
where $\ket{0}$ is the vacuum of the free theory, the subscript on $\hat Q_I$ means that this is an interaction-picture operator, and
\begin{equation}
    U_{\rm int}(\eta,\eta_0)=T\exp\left[-i\int_{\eta_0}^\eta\D \tilde\eta\, \hat H_{\rm int}(\tilde\eta) \right],
\end{equation}
with $T$ the time-ordering operator and $\hat H_{\rm int}$ the interaction Hamiltonian (expressed in the interaction picture).

We are interested in the one-loop contribution to the two-point correlation function of $\zeta$ (or $\gamma_{ij}$). The master formula in this case yields two terms, corresponding to two classes of diagrams: one with two insertions of the cubic-order Hamiltonian, $\hat{H}_3$,
\begin{equation} \label{eq:inin_cubic}
\begin{aligned}
    \braket{\hat{\zeta}_{\vec k}(\eta)\hat{\zeta}_{\vec p}(\eta)}^{\rm (1-loop)}&=2\,\mathrm{Re}\bigg[\int_{\eta_0}^{\eta}\D \eta_1\int_{\eta_0}^{\eta_1}\D \eta_2\langle0|\hat{H}_3(\eta_1)\hat{\zeta}_{\vec k}(\eta)\hat{\zeta}_{\vec p}(\eta)\hat{H}_3(\eta_2) \\
    &\quad -\hat{\zeta}_{\vec k}(\eta)\hat{\zeta}_{\vec p}(\eta)\hat{H}_3(\eta_1)\hat{H}_3(\eta_2)|0\rangle \bigg] \,,
    \end{aligned}
\end{equation}
and one with one insertion of the quartic-order Hamiltonian, $\hat{H}_4$,
\begin{equation}\label{eq:inin_quartic}
    \braket{\hat{\zeta}_{\vec k}(\eta)\hat{\zeta}_{\vec p}(\eta)}^{\rm (1-loop)}=-2\,\mathrm{Im}\left[
\int_{\eta_0}^\eta\mathrm d\eta_1 \braket{0|
\hat{H}_4(\eta_1)\hat\zeta_{\vec k}(\eta)\hat\zeta_{\vec p}(\eta)|0}
\right] \,.
\end{equation}

In order to lighten the notation, we will omit from now on the label `1-loop' and the hats in correlators. Using the usual notation where a prime on a correlator means that a factor of $(2\pi)^3$ and a momentum-conserving delta function have been stripped, we will moreover also frequently omit the labels indicating the external momenta $\vec k$ and $\vec p$ and external time $\eta$, i.e.
\begin{equation}
\braket{\hat{\zeta}_{\vec k}(\eta)\hat{\zeta}_{\vec p}(\eta)}^{\rm (1-loop)}\equiv (2\pi)^3\delta(\vec k+\vec p)\langle\zeta^2\rangle' \,,
\end{equation}
and we will quote results for the function $\langle\zeta^2\rangle'$ thus defined, or rather the dimensionless combination $\frac{k^3}{2\pi^2}\langle\zeta^2\rangle'$.

\subsection{Cutting off loop integrals}

Eqs.\ \eqref{eq:inin_cubic} and \eqref{eq:inin_quartic} are defined in terms of a reference initial time $\eta_0$. In standard set-ups with Bunch-Davies initial conditions, one is instructed to take $\eta_0\to-\infty$ (with an appropriate $i\epsilon$ prescription to ensure convergence) in order to correctly capture all contributions to the correlators. In models with imaginary speed of sound, on the other hand, one is dealing with an EFT with a finite range of validity: the theory is valid only after a time such that all modes have energies below the cutoff scale $\Lambda\equiv xH$, where $x$ is the parameter already introduced in Sec.\ \ref{subsec:quantization}. Given a comoving momentum $p$ and time $\eta$, this cutoff then translates into the inequality
\begin{equation} \label{eq:uv_cutoff}
-p|c_s|\eta\leq x \,,
\end{equation}
as a restriction on each mode for it to be in the regime of validity of the EFT. For tree-level computations, this implies that the time integrals in the in-in master formula are cut off by the time $\eta_0=-x/(k\cs)$, where $k$ is the largest among all external momenta. Additional care is needed with loop diagrams, since now one also must take into account the internal momenta running in the loops, as we discuss in detail next. Notice incidentally that \eqref{eq:uv_cutoff} is consistent with the regularization prescription discussed in \cite{Senatore:2009cf}, which emphasized the importance of setting a cutoff for \textit{physical} momenta, i.e.\ $p/a(\eta)\propto p\eta$, rather than for comoving momenta.

\subsubsection{Loop with one quartic vertex}

\begin{figure}
\centering
\includegraphics[width=0.5\linewidth]{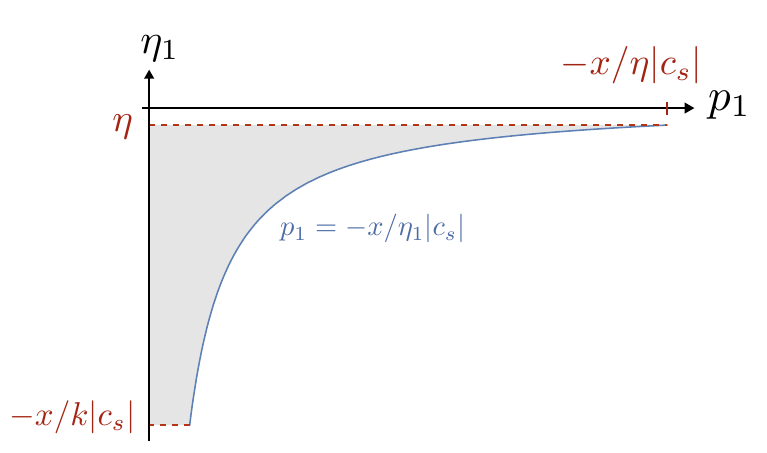}
\caption{The integration domain  in the loop diagram with one quartic vertex.}
\label{fig:domain_quartic}
\end{figure}

We start by scrutinizing \eqref{eq:inin_quartic}, which is the simplest case. The quartic Hamiltonian is first expressed in Fourier space, for example
\begin{equation}
H_{\zeta\zeta\zeta\zeta}^{(1)}=\mathscr D(2\pi)^3
\prod_{i=1}^4\int\frac{\D^3\vec p_i}{(2\pi)^3}
\delta\bigg(\sum_{j=1}^4\vec p_j\bigg)\zeta_{\vec p_1}'\zeta_{\vec p_2}'\zeta_{\vec p_3}'\zeta_{\vec p_4}' \,,
\end{equation}
so that, after using all delta functions (except for the one enforcing external momentum conservation), one is left with a single, one-dimensional momentum integral, i.e.\ the magnitude of the loop momentum (let us call it $p_1$), in addition to the time integral (let us call the integration variable $\eta_1$). All in all the loop contains four physical momenta (here $k$ is the external comoving momentum and $\eta$ is the external time): $k/a(\eta)$, $k/a(\eta_1)$, $p_1/a(\eta)$ and $p_1/a(\eta_1)$. These must all satisfy the bound \eqref{eq:uv_cutoff}, i.e.\ $-k\cs\eta\leq x$, $-k\cs\eta_1\leq x$, $-p_1\cs\eta\leq x$ and $-p_1\cs\eta_1\leq x$. The first and third of these are redundant since $\eta_1\leq \eta$; the second sets a lower limit on the time integral, while the last condition sets an upper limit on the momentum integral, that is
\begin{equation}
\int_{\eta_0}^\eta\mathrm d\eta_1\int_0^\infty \mathrm dp_1\quad\to\quad \int_{-x/(k|c_s|)}^\eta\mathrm d\eta_1\int_0^{-x/(\eta_1|c_s|)}\mathrm dp_1 \,.
\end{equation}
Since the remaining integral is manifestly convergent (the integrand does not contain any singularities), we are allowed to exchange the order of integration, which is often convenient both for analytical and numerical calculations. In this case the bound $-p_1\cs\eta\leq x$ sets the upper limit of the momentum integral,
\begin{equation}
\int_{-x/(k|c_s|)}^\eta\mathrm d\eta_1\int_0^{-x/(\eta_1|c_s|)}\mathrm dp_1 =\int_{0}^{-x/(|c_s|\eta)}\mathrm dp_1\int_{\eta_0}^{\eta}\mathrm d\eta_1 \,,
\end{equation}
with $\eta_0=\max\{-x/(k\cs),-x/(p_1\cs)\}$; see Fig.\ \ref{fig:domain_quartic}.

\subsubsection{Loop with two cubic vertices}

The loop integral in \eqref{eq:inin_cubic} contains two time integrations and a two-dimensional momentum integration, again after exploiting all available delta functions. A useful recasting of the momentum integrals is given by\footnote{To obtain this starting with the 6-dimensional momentum integral, one uses the identity
\begin{equation*}
k\int\D^3\vec p_1\int\D^3\vec p_2\,\delta^3(\vec p_1+\vec p_2+\vec k)f(p_1,p_2,k)=2\pi\int_0^\infty\D p_1\int_{|p_1-k|}^{p_1+k}\D p_2\,p_1p_2f(p_1,p_2,k) \,.
\end{equation*}
}
\begin{equation}
\int^\eta_{\eta_0}\D\eta_1\int_{\eta_0}^{\eta_1}\D\eta_2\int_0^\infty\D p_1\int_{|k-p_1|}^{k+p_1}\D p_2 \,.
\end{equation}
This integral must be cut off according to the prescription \eqref{eq:uv_cutoff}. 
The loop diagram contains a total of seven physical momenta, giving a total of seven constraints:
\begin{equation}
\begin{aligned}
&-q_i\cs\eta_j\leq x \,, \quad q_i\in\{k,p_1,p_2\} \,, \quad \eta_j\in\{\eta_1,\eta_2\} \,,\\
&\mbox{and}\quad -k\cs\eta\leq x \,.
\end{aligned}
\end{equation}
In particular, since $\eta_2<\eta_1$, we have $p_{1,2}\in (0,-x/\cs\eta_2)$. We may then identify two cases depending on the value of $\eta_2$.
\begin{itemize}
\item If $k<-x/(\cs\eta_2)<2k\Leftrightarrow -x/ (k\cs)<\eta_2<-x/(2k\cs)$, the integral may be divided into the following three parts:
\begin{equation}
\begin{aligned}
\left(
 \int \D p_1\int\D p_2
\right)_{1}&\equiv 
\int_0^{-x/(\cs\eta_2)-k} \D p_1\int_{k-p_1}^{k+p_1}\D p_2+\int_{-x/(\cs\eta_2)-k}^k\D p_1\int_{k-p_1}^{-x/(\cs\eta_2)}\D p_2 \\
&\quad +\int_{k}^{-x/(\cs\eta_2)}\D p_1\int_{p_1-k}^{-x/(\cs\eta_2)}\D p_2 \,;
\end{aligned}
\end{equation}

\item If $-x/(\cs\eta_2)>2k\Leftrightarrow \eta_2>-x/(2k\cs)$, the integral may be divided into the following three parts:
\begin{equation}
\begin{aligned}
\left(
 \int \D p_1\int\D p_2
\right)_{2}&\equiv 
\int_0^k \D p_1\int_{k-p_1}^{k+p_1}\D p_2+\int_k^{-x/(\cs\eta_2)-k}\D p_1\int_{p_1-k}^{p_1+k}\D p_2 \\
&\quad +\int_{-x/(\cs\eta_2)-k}^{-x/(\cs\eta_2)}\D p_1\int_{p_1-k}^{-x/(\cs\eta_2)} \D p_2 \,.
\end{aligned}
\end{equation} 
\end{itemize}
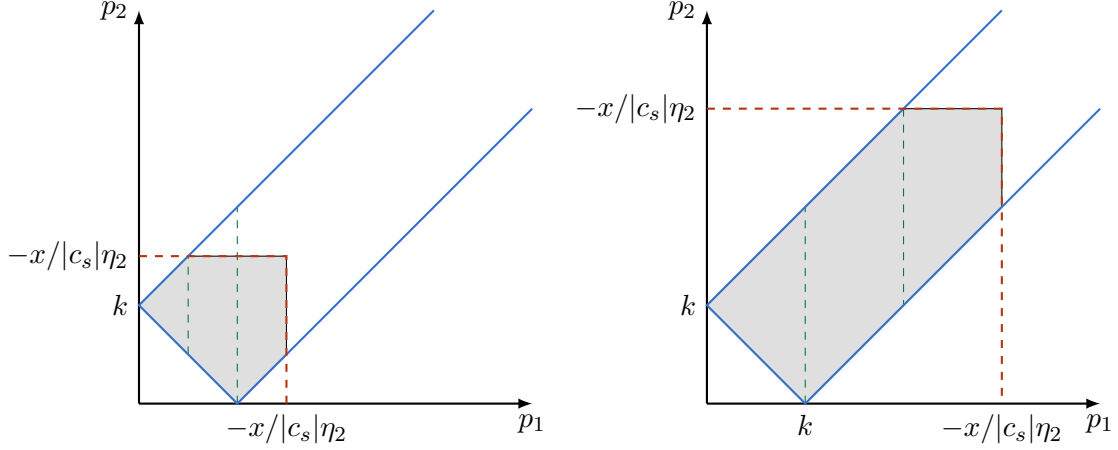
\begin{figure}
\centering
\begin{tikzpicture}[scale=1.3]
\draw[-latex,thick] (0,0)--(4,0)node[below]{$p_1$};
\draw[-latex,thick] (0,0)-- (0,4)node[left]{$p_2$};
\draw[fill=lightgray!50] (0,1)--(0.5,1.5)--(1.5,1.5)--(1.5,0.5)--(1,0)--(0,1);
\draw[LouisColor1,dashed] (1,0)--(1,2);
\draw[LouisColor1,dashed] (0.5,0.5)--(0.5,1.5);
\draw[thick,LouisBlue] (3,4)--(0,1) node[left,black]{$k$} -- (1,0)--(4,3);
\draw[thick,dashed,LouisOrange] (0,1.5)node[left,black]{$-x/\cs\eta_2$}--(1.5,1.5)--(1.5,0)node[below,black]{$-x/\cs\eta_2$};
\end{tikzpicture}
\begin{tikzpicture}[scale=1.3]
\draw[-latex,thick] (0,0)--(4,0)node[below]{$p_1$};
\draw[-latex,thick] (0,0)-- (0,4)node[left]{$p_2$};
\draw[fill=lightgray!50] (1,0)--(3,2)--(3,3)--(2,3)--(0,1)--(1,0);
\draw[LouisColor1,dashed] (1,0)--(1,2);
\draw[LouisColor1,dashed] (2,3)--(2,1);
\draw[thick,LouisBlue] (3,4)--(0,1) node[left,black]{$k$} -- (1,0)node[below,black]{$k$}--(4,3);
\draw[thick,dashed,LouisOrange] (0,3)node[left,black]{$-x/\cs\eta_2$}--(3,3)--(3,0)node[below,black]{$-x/\cs\eta_2$};
\end{tikzpicture}

\caption{The integration domain for the momentum integrals in the loop diagram with two cubic vertices. The gray region corresponds to the domain after introducing the EFT cutoff, which may be divided into three parts (indicated by the green dashed lines) for the purpose of calculating the integrals. The two graphs show the two qualitatively distinct cases arising from this division.}
\label{fig:int_area}
\end{figure}
The two cases are illustrated in Fig.\ \ref{fig:int_area}. Since $\eta_2$ is bounded by $\eta_1$, we further decompose the $\eta_1$ integral as follows:
\begin{equation}
\int_{-x/(k\cs)}^\eta \D \eta_1\equiv \left(\int\D\eta_1\right)_1+\left(\int\D\eta_1\right)_2 \,,
\end{equation}
with
\begin{equation}
\left(\int\D\eta_1\right)_1\equiv \int_{-x/(k\cs)}^{-x/(2k\cs)}\D\eta_1 \,,\qquad \left(\int\D\eta_1\right)_2\equiv \int_{-x/(2k\cs)}^\eta\D\eta_1 \,.
\end{equation}
We have assumed that the external time, $\eta$, is late enough so that $\eta>-x/(2k\cs)$. If $-x/(k\cs)<\eta_1<-x/(2k\cs)$, then we do not need to split the $\eta_2$-integral and we simply write
\begin{equation}
\int_{-x/k\cs}^{\eta_1}\D\eta_2\equiv \left(\int\D\eta_2\right)_1 \,,
\end{equation}
and the corresponding momentum integrals in this case are given by $(\int\D p_1\D p_2)_1$. If, on the other hand, $-x/(2k\cs)<\eta_1<\eta$ then we separate the integral as
\begin{equation}
\int_{-x/(k\cs)}^{\eta_1}\D\eta_2\equiv \left(\int\D\eta_2\right)_2+\left(\int\D\eta_2\right)_3 \,,
\end{equation}
with
\begin{equation}
\left(\int\D\eta_2\right)_2\equiv \int_{-x/(k\cs)}^{-x/(2k\cs)}\D \eta_2 \,,\qquad \left(\int\D\eta_2\right)_3\equiv \int_{-x/(2k\cs)}^{\eta_1}\D\eta_2 \,.
\end{equation}
In the first case among these two the associated momentum integral is $(\int\D p_1\D p_2)_1$, while it is $(\int\D p_1\D p_2)_2$ in the latter. All in all, the complete loop integral may be decomposed in terms of these definitions as
\begin{equation}
\boxed{
\begin{aligned}
\mathcal I&=\left(
\int\D\eta_1
\right)_1\left(\int\D\eta_2\right)_1\left(\int\D p_1\D p_2\right)_1\\
&\quad +\left(\int\D\eta_1\right)_2\left[\left(\int\D\eta_2\right)_2\left(
\int\D p_1\int\D p_2
\right)_1
+\left(\int\D\eta_2\right)_3\left(\int\D p_1\D p_2\right)_2
\right] \,.
\end{aligned}
}
\end{equation}

This result gives a correct definition of the loop integral in accordance with the cutoff prescription we have assumed. However, as we already mentioned, for practical purposes it is useful to exchange the order of integration. Performing the time integrals first, we have the limits of integration as given by
\begin{equation} \label{eq:exchange order integrals}
\eta_0\equiv \max\left\{
-\frac{x}{k\cs},-\frac{x}{p_1\cs},-\frac{x}{p_2\cs}
\right\}<\eta_2<\eta_1<\eta \,,
\end{equation}
and the upper bound on each momentum integral is $-x/(\cs\eta)$.

\begin{figure}
\centering
\begin{tikzpicture}[scale=1.5]
\draw[fill=lightgray!50] (0,-1)--(0,1)--(5.5,1)--(6.5,0)--(5.5,-1)--(0,-1);
\draw[dashed,LouisOrange] (5.5,-1.2)node[below]{$-\frac{2x}{\cs\eta}-2k$}--(5.5,1.2);
\draw[LouisOrange](6.5,-0.5)node[]{$-\frac{2x}{\cs\eta}-k$};
\draw[dashed,LouisOrange] (0,-1)--(1,0)node[below right]{$k$}--(0,1);
\draw[-latex,thick] (-0.5,0)--(7,0)node[below]{$t$};
\draw[-latex,thick] (0,-2)-- (0,2)node[left]{$s$};
\draw[](0.4,0.25) node[]{$A$};
\draw[](3.25,0.25) node[]{$C$};
\draw[](0.75,-0.75) node[]{$B$};
\draw[](0.75,0.75) node[]{$B$};
\draw[](5.85,0.25) node[]{$D$};
\draw[dashed,LouisOrange](1,-1.2)--(1,1.2);
\end{tikzpicture}
\caption{The integration domain for the momentum integrals in the loop diagram with two cubic vertices, in the $(t,s)$ plane (cf.\ \eqref{eq:ts variables}). The gray region corresponds to the domain after introducing the EFT cutoff, which may be divided into four parts ($A$, $B$, $C$ and $D$) for the purpose of calculating the integrals (see the main text for the rationale behind this division).
}
\label{fig:switch_region}
\end{figure}
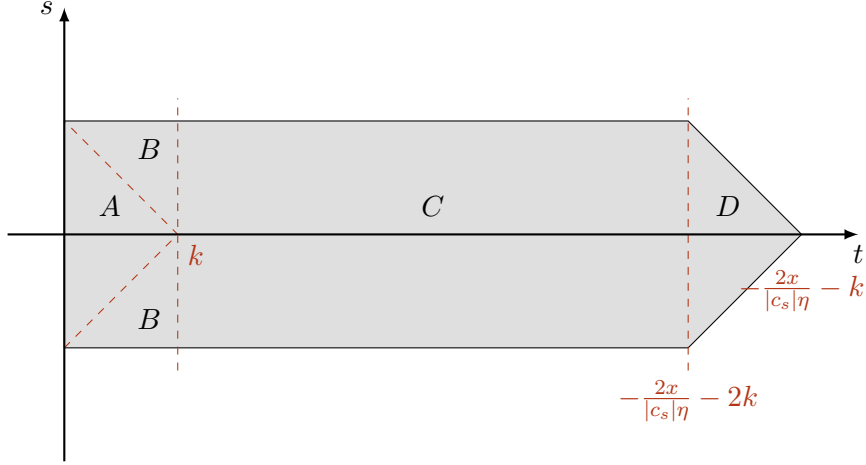

Both for practical convenience and to better understand the structure of the momentum integrals, we introduce the variables
\begin{equation} \label{eq:ts variables}
t\equiv p_1+p_2-k \,,\qquad s\equiv p_1-p_2 \,,
\end{equation}
and note that the Jacobian of the transformation is $1/2$, i.e.\ $\int\D p_1\D p_2=\frac{1}{2}\int\D t\D s$. Notice that $-k<s<k$ and $0<t<-2x/(\cs\eta)-k$, however the integration domain is not rectangular, because $t+s<-2x/(\cs\eta)-k$ and $t-s<-2x/(\cs\eta)-k$; see Fig.\ \ref{fig:switch_region}.

The domain may be divided into three main parts depending on $\eta_0$ (cf.\ \eqref{eq:exchange order integrals}). 

\begin{itemize}
\item Region $A$ is defined by the case $\eta_0=-x/(k\cs)$, equivalently $p_1,p_2<k$, i.e.\ $t+s<k$ and $t-s<k$. The loop integral in this region is given by
\begin{equation}
    \mathcal I_A=\frac12 \int_0^k\D t\int_{t-k}^{k-t}\D s\int_{-x/(k\cs)}^\eta\D\eta_1\int_{-x/(k\cs)}^{\eta_1}\D \eta_2 \,.
\end{equation}

\item Region $B$ has $p_1+p_2<2k$, with either $p_1>k$ or $p_2>k$. Equivalently, $t<k$, and $t+s>k$ or $t-s>k$. The resulting loop integral is given by
\begin{equation}
\begin{aligned}
\mathcal I_{B}&=\frac12 \int_0^k\D t\bigg[
\int_{-k}^{t-k}\D s\int_{-x/(p_2\cs)}^\eta\D\eta_1\int_{-x/(p_2\cs)}^{\eta_1}\D\eta_2 \\
&\quad+\int_{k-t}^k\D s\int_{-x/(p_1\cs)}^{\eta}\D \eta_1\int_{-x/(p_1\cs)}^{\eta_1}\D\eta_2\bigg] \,.
\end{aligned}
\end{equation}

\item Region $C$ has $2k<p_1+p_2<-2x/(\cs\eta)-k$, with either $p_1>k$ or $p_2>k$, or both $p_1,p_2>k$. Equivalently, $k<t<-2x/(\cs\eta)-2k$, and $\eta_0$ is determined by the sign of $s$. The corresponding loop integral is given by
\begin{equation}
\begin{aligned}
\mathcal I_{C}&=\frac12\int_k^{-2x/(\eta\cs)-2k}\D t\bigg[\int_{0}^k\D s\int_{-x/(p_1\cs)}^\eta\D \eta_1\int_{-x/(p_1\cs)}^{\eta_1}\D\eta_2 \\
&\quad+\int_{-k}^0\D s\int_{-x/(p_2\cs)}^\eta\D\eta_1\int_{-x/(p_2\cs)}^{\eta_1}\D \eta_2\bigg] \,.
\end{aligned}
\end{equation}

\item Region $D$ has $p_1+p_2>-2x/(\cs\eta)-k$, with $p_1,p_2<-x/(\cs\eta)$ as per the EFT bound, and again $\eta_0$ is determined by the sign of $s$. The loop integral in this domain reads
\begin{equation}
\begin{aligned}
\mathcal I_D&=\frac12\int_{-2x/(\cs\eta)-2k}^{-2x/(\cs\eta)-k}\D t\bigg[\int_{0}^{-2x/(\cs\eta)-k-t}\D s\int_{-x/(p_1\cs)}^{\eta}\D\eta_1\int_{-x/(p_1\cs)}^{\eta_1}\D \eta_2 \\
&\quad+\int_{2x/(\cs\eta)+k+t}^0\D s\int_{-x/(p_2\cs)}^\eta\D \eta_1\int_{-x/(p_2\cs)}^{\eta_1}\D \eta_2
\bigg] \,.
\end{aligned}
\end{equation}

\end{itemize}

%%%%%%%%%%%%%%%%%%%%%%
%%%%%%%%%%%%%%%%%%%%%%

\section{One-loop scalar power spectrum}
\label{sec:one loop scalar spectrum}

\subsection{Loops with two cubic vertices}

Having discussed the necessary calculational machinery, we are now ready to compute scalar one-loop diagrams in the in-in formalism. In this subsection we consider diagrams with two insertions of the cubic Hamiltonians $H^{(1)}_{\zeta\zeta\zeta}$ and $H^{(2)}_{\zeta\zeta\zeta}$ displayed in \eqref{eq:cubicHam} and \eqref{eq:cubicHam2}.

\subsubsection*{Two $H_{\zeta\zeta\zeta}^{(1)}$-vertices.}

\begin{figure}
\centering
\includegraphics{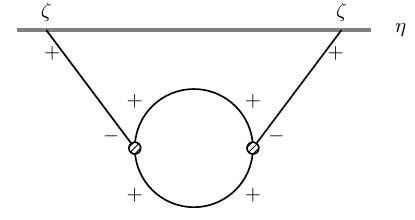}
\includegraphics{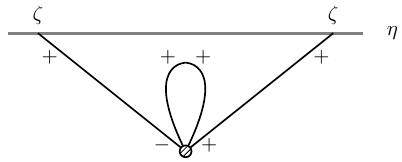}
\caption{Leading scalar one-loop diagrams in the large-$x$ approximation.
The sign $\pm$ associated with each mode indicates if it is growing or decaying, see Eq.~\eqref{eq:growing-decaying modes}.}
\label{fig:feyndiags1}
\end{figure}

We substitute $H_{\zeta\zeta\zeta}^{(1)}$ into the in-in formula \eqref{eq:inin_cubic} and carry out the Wick contractions (36 in total, all equivalent) to eventually obtain\footnote{Notice that we do not include contractions corresponding to tadpole diagrams (cf.\ Fig.\ \ref{fig:feyndiags1}). Indeed, contributions from tadpole diagrams vanish identically for purely derivative interactions~\cite{Senatore:2009cf}, as considered in this paper. The simple reason is that a tadpole diagram includes a propagator with zero momentum, while time and spatial derivatives of the mode function vanish at zero momentum.}
\begin{equation}
\begin{aligned}
\braket{\zeta_{\vec k}(\eta)\zeta_{\vec p}(\eta)}&=-\frac{144 \mathscr C^2}{H^{2}}\delta^3(\vec k+\vec p)\int_{\eta_{0}}^{\eta} \mathrm{d}\eta_1 \eta_1^{-1}\int_{\eta_{0}}^{\eta_{1}}  \mathrm{d}\eta_2 \eta_2^{-1}\int \mathrm{d}^3 \vec p_1 \mathrm{d}^3 \vec p_2\,\delta^3(\vec p_1+\vec p_2-\vec k) \\
&\quad\times\mathrm{Im}\left[\zeta_{k}'(\eta_1)\zeta_{k}^*(\eta)\right]\mathrm{Im}\left[\zeta_{k}(\eta)\zeta_{k}^{*\prime}(\eta_2)\zeta_{p_1}^{\prime}(\eta_1)\zeta_{p_1}^{*\prime}(\eta_2)\zeta_{p_2}'(\eta_1)\zeta_{p_2}^{*\prime}(\eta_2)\right] \,.
\end{aligned}
\end{equation}
So far no approximation has been made. Next we consider the dominant contribution in the large-$x$ expansion, which we find to correspond to the case when the two \textit{internal} modes carrying the external momentum $\vec k$ are decaying, i.e.\ $\zeta_{k}(\eta_1)=\zeta_{k,-}(\eta_1)$ and $\zeta_{k}(\eta_2)=\zeta_{k,-}(\eta_2)$, with all other fluctuations being growing modes; see Fig.\ \ref{fig:feyndiags1} for an illustration. Notice that indeed at least two modes must be decaying, for otherwise the imaginary parts in the previous equation would evaluate to zero. Certainly, there are several other choices for the decaying modes, but we have found through an explicit calculation\footnote{Although our analytical results are restricted to the leading order contributions for each type of loop diagram, in every case we have checked through numerical computations that the neglected terms are indeed subleading.} that the aforementioned choice gives the leading terms, namely ones proportional to $x^5$, $x^4$ and $x^3$ in the final normalized result (see below), while other choices contribute at most at order $x^2$.

Focusing on this choice, we then arrive at the following expression for the one-loop dimensionless power spectrum:
\begin{equation}
\frac{k^3}{2\pi^2}\braket{\zeta^2}'=\frac{144\cs^{12}\mathscr C^2}{\pi H^2}A^6\mathrm{Im}[B]^2 e^{2k\cs\eta}(k\cs\eta-1)^2 \mathscr I_{\zeta\zeta\zeta}^{(1)} \,,
\end{equation}
in terms of the integral
\begin{equation} \label{eq:I1 integral}
\begin{aligned}
\mathscr I^{(1)}_{\zeta\zeta\zeta}&=\int_{\eta_0}^\eta\D\eta_1\int_{\eta_0}^{\eta_1}\D\eta_2\int_0^\infty \D p_1\int_{|p_1-k|}^{p_1+k}\D p_2\, \eta_1^2\eta_2^2p_1^2p_2^2 e^{(p_1+p_2-k)\cs(\eta_1+\eta_2)} \\
&=\frac{1}{32}
\int_{\eta_0}^\eta\D\eta_1\int_{\eta_0}^{\eta_1}\D\eta_2\int_0^\infty\D t\int_{-k}^k\D s\, \eta_1^2\eta_2^2 \left(
\left(t+k\right)^2-s^2
\right)^2e^{t\cs(\eta_1+\eta_2)} \,.
\end{aligned}
\end{equation}

Next we regularize and manipulate this integral following the steps explained in the previous section. Before quoting the exact, analytical result, let us note that the dominant scaling with $x$ may be inferred without carrying out the computation explicitly. First of all, the exponent of $e^{t\cs(\eta_1+\eta_2)}$ sets the dominant scale $-t\cs(\eta_1+\eta_2)\sim1\Rightarrow t\sim \frac{-1}{\cs(\eta_1+\eta_2)}\lesssim -2x/(\cs\eta_{1,2})$, where $-2x/(\cs\eta_{1,2})$ is roughly the upper bound on $t$. Then $x$ enters in the result only through the lower limit $\eta_0$ of the time integrals.
Consider first the contribution from region $A$ (cf.\ Fig.\ \ref{fig:switch_region}), where the integrals are dominated by $\eta_1\sim \eta_2\sim -x/(k\cs)$. This leads to $t\sim k/x\ll k$, that is, the dominant contribution happens at the limit $p_1+p_2-k\sim k/x\ll k$.
Setting $t\sim k/x$ and $\eta_1\sim\eta_2\sim -x/(k\cs)$, one immediately sees that the integral scales as $\mathscr I_{\zeta\zeta\zeta,A}^{(1)}\propto x^5$ (noting that the integration measures $\int\D t$ and $\int\D\eta_{1,2}$ contribute factors of order $k/x$ and $-x/k\cs$, respectively).
On the other hand, in regions $B$, $C$ and $D$, we would have different dominant scales, namely $\eta_1,\eta_2\sim -x/(p_{1,2}\cs)$, i.e.\ $t\sim p_{1,2}/x\ll k$ and $s\sim k$. These parts of the integral thus scale as $x^2$ and are therefore subdominant. We conclude then that the leading contribution to the integral comes from the domain $A$: $\mathscr I_{\zeta\zeta\zeta}^{(1)}= \mathscr I_{\zeta\zeta\zeta,A}^{(1)}+\mathcal{O}(x^2)$.

An explicit calculation confirms this quick estimate. Recalling the expressions for the parameters $A$, $B$ and $\mathscr C$, as well as the dimensionless tree-level power spectrum $\mathcal{P}_\zeta$, we eventually get
\begin{equation} \label{eq:1loop_cubic}
\boxed{
\begin{aligned}
\frac{k^3}{2\pi^2}\braket{\zeta^2}'&=\mathcal P_\zeta^2\mathcal{A}^2 \left(
\frac{1}{|c_s|^2}+1
\right)^2
e^{2k|c_s|\eta}\left(
k|c_s|\eta-1
\right)^2 \\
&\quad \times\left[
\frac{3(2\log 2-1)}{800}x^5+\frac{3(3-4\log 2)}{128}x^4+
\frac{8\log 2-5}{96}x^3+\mathcal O(x^2)
\right] \,.
\end{aligned}
}
\end{equation}
Remarkably, the integral $\mathscr I_{\zeta\zeta\zeta}^{(1)}$ is time-independent at the orders shown in the large-$x$ expansion, although we emphasize that subdominant terms, starting at $\mathcal{O}(x^2)$, do depend on the external time $\eta$.

\begin{figure}
\centering
\includegraphics[scale=0.7]{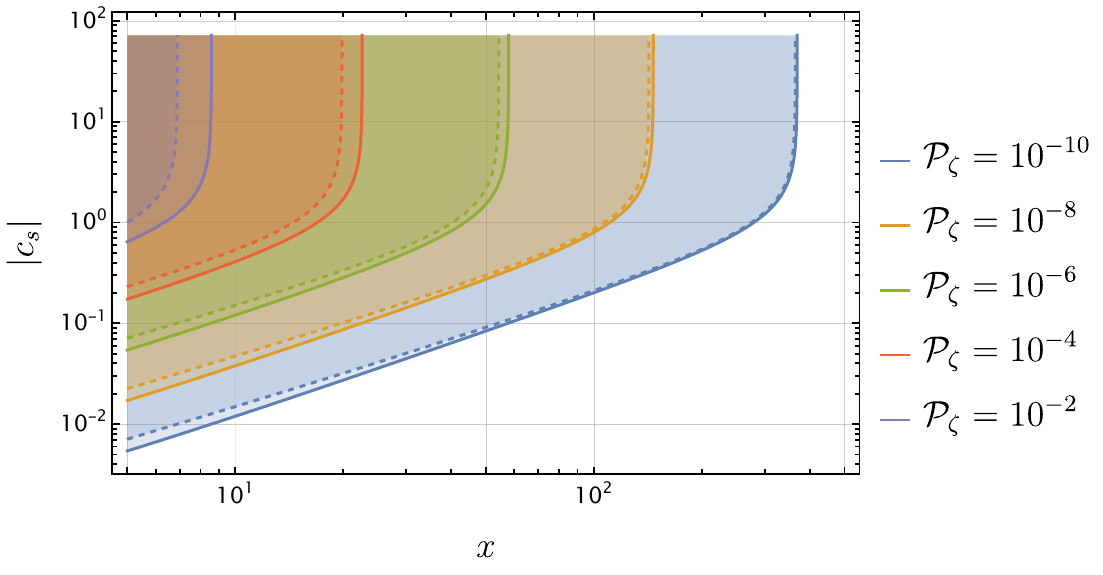}
\caption{Consistency with perturbativity as dictated by the criterion $\mathcal{P}_{\zeta}^{\rm(1-loop)}/\mathcal{P}_{\zeta}^{\rm(tree)}<1$ (colored regions) as function of $x$ and $\cs$, for several values of $\mathcal P_\zeta$ (the tree-level dimensionless power spectrum). \textit{Solid lines:} $\mathcal{P}_{\zeta}^{\rm(1-loop)}$ as given by Eq.\ \eqref{eq:1loop_cubic} with $\mathcal{A}=1$, i.e.\ the case of the diagram with two insertions of $H_{\zeta\zeta\zeta}^{(1)}$. \textit{Dashed lines:} The same in the case of the loop diagram with two insertions of $H_{\zeta\zeta\zeta}^{(2)}$, cf.\ Eq.\ \eqref{eq:1loop_cubic2}. Note that in both cases the late-time limit $\eta\to0^-$ has been taken.
}
\label{fig:PB}
\end{figure}

Defining $\mathcal{P}_{\zeta}^{\rm(1-loop)}\equiv \lim_{\eta\to0^-}\frac{k^3}{2\pi^2}\braket{\zeta^2}'$, one finds a simple expression for the ratio $\mathcal{P}_{\zeta}^{\rm(1-loop)}/\mathcal{P}_{\zeta}^{\rm(tree)}$ that we use to diagnose the breakdown of perturbativity. Taking $\mathcal{A}\sim1$ one has
\begin{equation}
\frac{\mathcal{P}_{\zeta}^{\rm(1-loop)}}{\mathcal{P}_{\zeta}^{\rm(tree)}}\sim \mathcal{P}_\zeta\left(
\frac{1}{|c_s|^2}+1
\right)^2 x^5 \,.
\end{equation}
We see that this can be dangerously large already for moderate values of $x$. For example, using the CMB value $\mathcal{P}_\zeta\sim10^{-9}$ and $\cs\sim 10^{-1}$, one has $\mathcal{P}_{\zeta}^{\rm(1-loop)}/\mathcal{P}_{\zeta}^{\rm(tree)}\sim1$ for $x\sim10$. On the positive side, models that predict $\cs=\mathcal{O}(1)$, such as hyper-inflation \cite{Brown:2017osf,Fumagalli:2019noh}, appear to be consistent even for values of $\mathcal{P}_\zeta$ that are strongly enhanced relative to CMB scales. A more precise illustration is provided in Fig.\ \ref{fig:PB}, where we show the ``exclusion'' region $\mathcal{P}_{\zeta}^{\rm(1-loop)}/\mathcal{P}_{\zeta}^{\rm(tree)}>1$ as function of $x$ and $\cs$ for several values of ${\cal P}_\zeta$.\footnote{Unlike in the standard situation with real $c_s$, here we have no fundamental obstruction for allowing $\cs>1$. Not that this helps too much with the issue of perturbativity, since for large $\cs$ the bound then becomes degenerate in $\cs$. \label{footnote:large cs}}

\subsubsection*{Two $H_{\zeta\zeta\zeta}^{(2)}$-vertices.}

The second contribution we consider is the loop diagram with two insertions of the $H_{\zeta\zeta\zeta}^{(2)}$ Hamiltonian in Eq.\ \eqref{eq:cubicHam2}. The calculation follows precisely the same steps as in the previous paragraph, the only additional complication being that there are now several inequivalent permutations. We provide details of the computation in Appendix \ref{app:H2 scalar loop}, while here we only quote the final result:
\begin{equation} \label{eq:1loop_cubic2}
\boxed{
\begin{aligned}
\frac{k^3}{2\pi^2}\braket{\zeta^2}'&=\mathcal P_\zeta^2\left(
\frac{1}{|c_s|^2}+1
\right)^2
e^{2k|c_s|\eta}\left(
k|c_s|\eta-1
\right)^2 \\
&\quad \times\left[
\frac{3(2\log 2-1)}{800}x^5+\frac{4\log 2+5}{640}x^4+
\frac{16\log 2-19}{480}x^3+\mathcal O(x^2)
\right] \,.
\end{aligned}
}
\end{equation}

We remark on the close similarity of this result with the case of two $H_{\zeta\zeta\zeta}^{(1)}$-vertices, Eq.\ \eqref{eq:1loop_cubic}. In particular, both feature the same $x^5$ dominance in the large-$x$ approximation, which indeed can be explained through a simple scaling argument; see the next paragraph. We also observe the same $1/\cs^4$ enhancement in the limit of small $\cs$, consistent with our claim that the perturbativity bound is not qualitatively affected by considering all types of vertices. This is further illustrated in Fig.\ \ref{fig:PB} displaying the exclusion regions as functions of $x$, $\cs$ and $\mathcal{P}_\zeta$.

\subsubsection*{Two mixed vertices.}

We have not carried out the calculation of the loop diagram with one insertion of $H_{\zeta\zeta\zeta}^{(1)}$ and one insertion of $H_{\zeta\zeta\zeta}^{(2)}$, a cumbersome task due to the large number of inequivalent permutations. Nevertheless, from the previous results one can clearly anticipate a qualitatively identical outcome, so that the full result will not be affected, in order of magnitude, from our neglecting this contribution.

Indeed, the scaling argument discussed previously applies generally to all the cubic vertices in the EFT of inflation in the regime we are considering. The only novelty here, compared to the easiest case with two insertions of $H_{\zeta\zeta\zeta}^{(1)}$, is that the $H_{\zeta\zeta\zeta}^{(2)}$ vertex carries both time and space derivatives of the mode function. Notice however that, as far as the exponential term in the integral is concerned (cf.\ \eqref{eq:I1 integral}), time and space derivatives contribute equally, so the dominant scale is still set by $t\sim \frac{-1}{\cs(\eta_1+\eta_2)}$. Powers of internal momenta therefore do not yield additional powers of $x$ in this subdomain of integration. This explains why both of our explicit results, Eqs.\ \eqref{eq:1loop_cubic} and \eqref{eq:1loop_cubic2}, feature the same large-$x$ scaling, and we similarly conclude that the terms we have neglected must also enjoy the same $x^5$ enhancement.

%%%%%%%%%%%%%%%%%%%%%%%%%%%%%

\subsection{Loops with one quartic vertex}

We move on to study one-loop diagrams with a single quartic vertex. Recall from Sec.\ \ref{sec:preliminary} that we will consider for simplicity only the contribution from the $\zeta^{\prime4}$ vertex given in \eqref{eq:quarticHam}, since neglecting other terms will not affect the result qualitatively, in particular concerning the large-$x$ behavior; see Appendix \ref{app:other quartic} for further explanations as well as the explicit result for the $(\partial\zeta)^4$ vertex.

The relevant in-in integral is given in \eqref{eq:inin_quartic}, into which we substitute $H_{\zeta\zeta\zeta\zeta}^{(1)}$ to produce
\begin{equation}
\langle\zeta_{\vec k}(\eta)\zeta_{\vec p}(\eta)\rangle=-24\mathscr D\delta^3(\vec k+\vec p)\int_{\eta_0}^\eta\mathrm d\eta_1\int\mathrm d^3\vec p_1\,\mathrm{Im}\left[
\zeta'_k(\eta_1)\zeta_k^*(\eta)\zeta_k'(\eta_1)\zeta_k^*(\eta)\zeta_{p_1}'(\eta_1)\zeta_{p_1}'^*(\eta_1)
\right] \,,
\end{equation}
after performing the Wick contractions and taking into account the 12 equivalent permutations. In order to isolate the leading behavior in the large-$x$ approximation, it is clear that now it suffices to take only one of the modes to be decaying. We find that choosing either of the internal modes carrying the external momentum $\vec k$ yields the dominant contribution, i.e.\ $\zeta_k(\eta_1)=\zeta_{k,-}(\eta_1)$, just like in the case of the loop diagrams with two cubic vertices; this is also illustrated in Fig.\ \ref{fig:feyndiags1}. Eventually we obtain the following result for the contribution of this loop diagram to the dimensionless power spectrum:
\begin{equation}
\frac{k^3}{2\pi^2}\langle\zeta^2\rangle'=-\frac{96\cs^8\mathscr{D}}{\pi}A^5\mathrm{Im}[B]ke^{2k\cs\eta}(k\cs\eta-1)^2\mathscr I_{\zeta\zeta\zeta\zeta} \,,
\end{equation}
where
\begin{equation}
\mathscr I_{\zeta\zeta\zeta\zeta}=\int_{\eta_0}^{\eta}\D\eta_1\int_0^{\infty}\D p_1\, \eta_1^4p_1^3e^{2p_1\cs\eta_1} \,.
\end{equation}
This integral must be regularized following the prescription described in Sec.\ \ref{sec:preliminary}. As we have seen, this procedure introduces an explicit dependence on the cutoff parameter $x$, which is easy to estimate by means of the scaling argument we have already used. The exponential term $e^{2p_1\cs\eta_1}$ suppresses the integral except in the subdomain $-p_1\cs\eta_1= \mathcal O(1)$; in this region, then, the integrand is independent of $x$, so the time integral yields $\int_{-x/(k\cs)}^\eta \D\eta_1\,\mathcal{O}(x^0)\propto x$, i.e.\ a linear scaling with $x$ at leading order.

This quick estimate is borne out by an explicit computation of the integral:
\begin{equation} \label{eq:quaticLoop}
\boxed{
\frac{k^3}{2\pi^2}\braket{\zeta^2}'= -\mathcal P_\zeta^2\mathcal D e^{2k|c_s|\eta}\left(
k|c_s|\eta-1
\right)^2
\left[\frac{9x}{4|c_s|^4}+\mathcal O(x^0)\right] \,.
}
\end{equation}
We have only quoted the $\propto x$ term since at $\mathcal O(x^0)$ we also have contributions from other combinations of growing versus decaying modes that we neglected. We remark that \eqref{eq:quaticLoop} features the same time dependence of the diagrams we computed previously, Eqs.\ \eqref{eq:1loop_cubic} and \eqref{eq:1loop_cubic2}. We emphasize however that this only holds at leading order in $x$, i.e.\ the $\mathcal O(x^0)$ terms that we omitted in \eqref{eq:quaticLoop} do depend on time and we do not expect this agreement to uphold at subleading orders.

In the late-time limit, and assuming $\mathcal{D}\sim1$, we have the estimate
\begin{equation}
\frac{\mathcal{P}_{\zeta}^{\rm(1-loop)}}{\mathcal{P}_{\zeta}^{\rm(tree)}}\sim \mathcal P_\zeta\frac{x}{\cs^4} \,,
\end{equation}
for the effect of the one-loop power spectrum, for this particular vertex, relative to the tree-level result. We observe that this is subleading in comparison with the contributions from diagrams with two cubic vertices in the large-$x$ expansion, at least assuming $\cs$ is not too small. We show in Appendix \ref{app:other quartic} that other quartic vertices should have the same large-$x$ behavior and are therefore also negligible in this approximation.

%%%%%%%%%%%%%%%%%%%%%%%%%%%%%%%

\subsection{Counterterms and higher derivative corrections}

We make an aside here to comment on the question of whether the one-loop results we have computed are actually physical. In standard QFT one knows that, in cutoff regularization, powers of the cutoff are unphysical since they are absorbed by counterterms upon renormalization. It is then natural to ask if the same might happen in the EFT of inflation with imaginary speed of sound.

As a first approach to this question, we can compute the effect of higher-derivative corrections that could a priori serve as counterterms in the renormalization procedure. This has been done explicitly in Ref.\ \cite{Senatore:2009cf} in the context of the standard EFT of inflation. Here we focus for simplicity on a single higher-derivative operator:
\begin{equation} \label{eq:counterterm1}
S_{\rm ct}=\mathcal{K}\int\D\eta\D^3\vec x\,a^{-2}(\zeta''')^2 \,,
\end{equation}
where $\mathcal{K}$ is a constant. There are actually two other terms at this order in the derivative expansion~\cite{Senatore:2009cf}, however we have checked that they yield similar results and hence do not affect the reasoning that follows.

Treating \eqref{eq:counterterm1} as an interaction in the in-in formalism, we find its contribution to the scalar power spectrum to be
\begin{equation}
\braket{\zeta_{\vec k}(\eta)\zeta_{\vec p}(\eta)}_{\rm ct}=4\mathcal{K}(2\pi)^3\delta^3(\vec k+\vec p)\int_{\eta_0}^\eta\D\eta_1\,a^{-2}(\eta_1)\operatorname{Im}\left[\zeta_{k}'''(\eta_1)^2\zeta_k^*(\eta)^2\right] \,.
\end{equation}
Regularizing the integral and isolating the leading terms in the large-$x$ approximation, we eventually obtain
\begin{equation} \label{eq:counterterm2}
\begin{aligned}
\frac{k^3}{2\pi^2}\langle\zeta^2\rangle_{\rm ct}'&=16\pi^2\mathcal{K}H^2\mathcal{P}_{\zeta}^2e^{-2x}\cs^3(\rho\sin\psi)e^{2k|c_s|\eta}(k|c_s|\eta-1)^2 \\
&\quad\times\left[\frac{x^5}{5}-\frac{4x^3}{3}+\rho\cos\psi\left(\frac{x^4}{2}+x^3\right)+\mathcal{O}(x^2)\right] \,.
\end{aligned}
\end{equation}
To get this result we have again considered the case when the two external modes are growing, which one can easily verify to produce the greatest enhancement in powers of $x$.\footnote{Unlike in the loop computations, we have considered here also the next-to-leading order combination of growing versus decaying modes, namely the case where two internal modes are decaying, which notably enter here at $\mathcal{O}(x^4)$.}

A few remarks are in order. First, \eqref{eq:counterterm2} features the same dependence on the external time $\eta$ as the one-loop results, cf.\ \eqref{eq:1loop_cubic}, \eqref{eq:1loop_cubic2} and \eqref{eq:quaticLoop}. However, as already explained, this is simply a consequence of our focusing on the contributions where the two external modes are growing. Since by assumption the external time is taken to be late enough, the dependence on $\eta$ brought in by the time integral is necessarily subleading. So the agreement of \eqref{eq:counterterm2} and the one-loop effects with regards to the time dependence should not be taken as a sign that the latter are unphysical. One could, of course, attempt to cancel these one-loop corrections by appropriately choosing the constant $\mathcal{K}$, however this can never be done for all times $\eta$ and for all powers of $x$, so there is no reason why precisely the highest powers must be canceled via this procedure. In standard renormalization, \textit{all} power-law divergences must be canceled, precisely because they are unphysical. We thus conclude that powers of $x$ in our set-up are perfectly physical; in fact, they already appear at tree level \cite{Garcia-Saenz:2018vqf,Fumagalli:2019noh}. A second observation is that \eqref{eq:counterterm2} is actually exponentially suppressed in the large-$x$ approximation. This is both satisfactory, as it means that higher-derivative corrections do not contaminate our one-loop estimates, and also consistent with our claim that the latter effects are physical: in order for \eqref{eq:counterterm2} to cancel the one-loop results, the constant $\mathcal{K}$ must scale as $e^{2x}$. While formally nothing prevents one from making this choice, it would be again at odds with the standard logic of renormalization, where counterterms precisely absorb power-law divergences, but one does not encounter for instance exponentials of the cutoff.

%%%%%%%%%%%%%%%%%%%%%%%%%%%%%%%
%%%%%%%%%%%%%%%%%%%%%%%%%%%%%%%

\section{One-loop tensor power spectrum}
\label{sec:one loop tensor spectrum}

\begin{figure}
\centering
\includegraphics{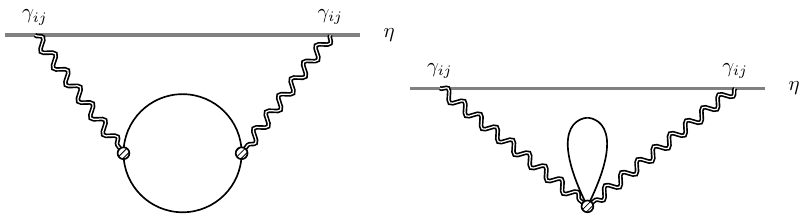}
\label{fig:tensor loops}
\caption{One-loop diagrams that contribute to the tensor power spectrum. Tensor modes are denoted by double-wiggly lines.
}
\end{figure}

In this Section we study the scalar one-loop corrections to the tensor power spectrum,\footnote{Loop corrections to the tensor power spectrum with real $c_s$ have been studied in several recent works \cite{Kong:2024lac,Ballesteros:2024qqx,Braglia:2025cee,Ema:2025ftj,Fang:2025hid}.} focusing on the vertices listed in \eqref{eq:cubicHamMix}, \eqref{eq:quarticHam tensor1} and \eqref{eq:quarticHam tensor2}. The two types of loop diagrams are shown in Fig.\ \ref{fig:tensor loops}. A key difference relative to the scalar spectrum case of the previous Section is that here we may take all $\zeta$ modes as growing, with a correspondingly dangerous enhancement in the large-$x$ regime.

\subsection{Loops with two cubic vertices}
\label{subsec:tensor cubic}

As explained, at the order we are working in the slow-roll and derivative expansions, there is a single mixed scalar-tensor cubic vertex, Eq.\ \eqref{eq:cubicHamMix}. The relevant in-in integral is given in \eqref{eq:inin_cubic}, of course replacing $\hat\zeta$ with $\hat\gamma$ in the external modes. Performing the Wick contractions we get
\begin{equation} \label{eq:1looptensor}
\begin{aligned}
\;&\braket{\gamma_{ij,\vec k}(\eta)\gamma_{ij,\vec p}(\eta)}=-16\mathscr E^2 P_{kl,mn}(\hat{\mathbf k})\delta^3(\mathbf k+\mathbf p) \\
&\quad\times\int_{-\infty}^\eta\D\eta_1\,a^2(\eta_1)\int_{-\infty}^{\eta_1}\D\eta_2\,a^2(\eta_2)\int\D^3\vec p_1\int\D^3\vec p_2\,\delta^3(\mathbf p_1+\mathbf p_2-\mathbf k)\\
&\quad\times\mathbf p_1^k\mathbf p_2^l\mathbf p_1^m\mathbf p_2^n\, \mathrm{Im}\left[\gamma_k(\eta_1)\gamma_k^*(\eta)\right]\mathrm{Im}\left[
\zeta_{p_1}(\eta_1)\zeta_{p_1}^*(\eta_2)\zeta_{p_2}(\eta_1)\zeta_{p_2}^*(\eta_2)\gamma_k(\eta)\gamma_k^*(\eta_2)\right] \,.
\end{aligned}
\end{equation}
Recall that the tensor mode functions are given by the Bunch-Davies state, while for the scalar fluctuations we isolate the leading behavior in $x$ by considering only growing modes. Thus we arrive at\footnote{Here we made use of the identity
\begin{equation}
P_{ij,kl}(\hat{\vec k})\vec p_1^i\vec p_2^j\vec p_1^k\vec p_2^l=\frac12\big[p_1^2-(\vec p_1\cdot \hat{\vec k})^2 \big]\big[p_2^2-(\vec p_2\cdot\hat{\vec k})^2\big]\,.
\end{equation}
}
\begin{equation} \label{eq:1looptensor2}
\begin{aligned}
\frac{k^3}{2\pi^2}\braket{\gamma^2}'&=-\frac{4\epsilon^2\mathcal{P}_\zeta^2}{k^2} \int_{-\infty}^{\eta}\frac{\D \eta_1}{\eta_1^2}\int_{-\infty}^{\eta_1}\frac{\D \eta_2}{\eta_2^2} e^{-ik(2\eta-\eta_1-\eta_2)} \\
&\quad\times\left(e^{2ik(\eta-\eta_1)}(i+k\eta)(-i+k\eta_1)-(-i+k\eta)(i+k\eta_1)\right) \\
&\quad\times\left(e^{2ik(\eta-\eta_2)}(i+k\eta)(-i+k\eta_2)-(-i+k\eta)(i+k\eta_2)\right) \\
&\quad\times\int_{0}^{\infty}\D u\int_{|1-u|}^{1+u}\D v \left[\frac{4v^2-(1+v^2-u^2)^2}{4uv}
\right]^2e^{(u+v)k|c_s|(\eta_1+\eta_2)} \\
&\quad\times (uk|c_s|\eta_1-1)(uk|c_s|\eta_2-1)(vk|c_s|\eta_1-1)(vk|c_s|\eta_2-1) \,,
\end{aligned}
\end{equation}
where we introduced the variables $u\equiv p_1/k$ and $v\equiv p_2/k$. Observe that time integrals here are not cut off because the momentum of the tensor modes can be arbitrarily large. On the other hand, the momentum integrals relate to the momenta of the scalar modes running in the loop, and are therefore bounded by $-x/\cs\eta_2$. Remarkably, however, a quick inspection of the integral reveals that the result is actually independent of $x$ at leading order in the large-$x$ expansion. Equivalently, the limit $x\to\infty$ converges and captures this leading-order behavior. To see this, notice that the exponential term suppresses the integral except in the region around $u+v\sim -\frac{1}{k\cs(\eta_1+\eta_2)}$, which is parametrically smaller than the upper bound on $u+v$, i.e.\ $-2x/(k\cs\eta_2)$. Put conversely, any $x$-dependence of the integral must arise from the upper limit of integration, near which the integrand is exponentially suppressed, $\exp[(u+v)k\cs(\eta_1+\eta_2)]\lesssim e^{-x}$.

\begin{figure}
\centering
\includegraphics[width=0.7\linewidth]{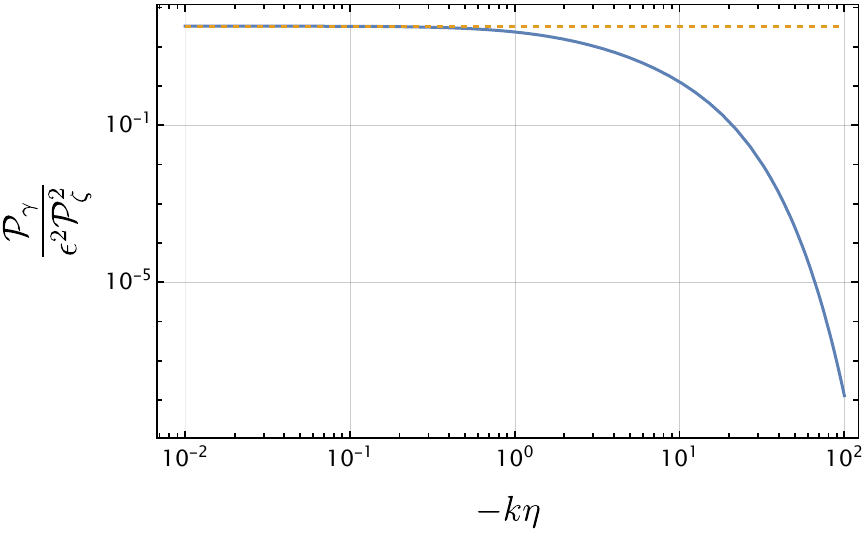}
\caption{Time evolution of the one-loop correction $\mathcal P_\gamma^{\rm (1-loop)}$ from the cubic vertex $H_{\gamma\zeta\zeta}$, at leading order in the large-$x$ approximation, i.e.\ Eq.\ \eqref{eq:1looptensor2}, using $\cs=10^{-1}$ and normalized by $\epsilon^2\mathcal P_\zeta^2$. The horizontal dashed line is the late-time limit given by Eq. \eqref{eq:tensor_inin_full}.
}
\label{fig:finitetime_tensor}
\end{figure}

The integral in \eqref{eq:1looptensor2} can be performed analytically for any finite external time $\eta$, although the result is long and not particularly illuminating. We display the time evolution in Fig.\ \ref{fig:finitetime_tensor}, which shows that the loop correction grows monotonically with time and quickly approaches a constant value after horizon crossing. Evaluating at $\eta\to0^-$ we obtain
\begin{equation} \label{eq:tensor_inin_full}
\begin{aligned}
\mathcal{P}_\gamma^{\rm(1-loop)}&=\frac{\epsilon^2\mathcal P_\zeta^2}{45}\bigg[-\frac{15|c_s|^4+364|c_s|^2+357}{1+\cs^2} \\
&\quad+\frac{3}{\cs}\left(\frac{\pi}{2}-\arctan\cs\right)\left(5 |c_s|^4+118|c_s|^2+41\right)+\mathcal{O}(x^{-1})\bigg] \,.
\end{aligned}
\end{equation}

\begin{figure}
\centering
\includegraphics[scale=0.7]{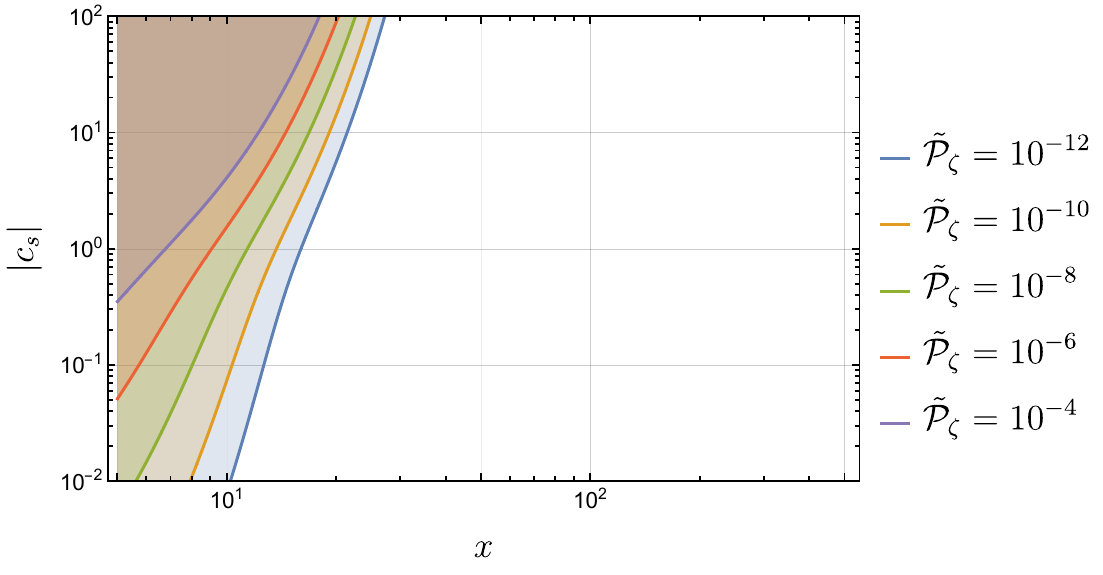}
\caption{
Consistency with perturbativity as dictated by the criterion $\mathcal{P}_\gamma^{\rm(1-loop)}/\mathcal{P}_\gamma^{\rm(tree)}<1$ (colored regions) as function of $x$ and $\cs$, for several values of $\tilde{\mathcal P}_{\zeta}\equiv {\epsilon\mathcal P_\zeta}/{(\rho\sin \psi)}$, with $\mathcal{P}_{\gamma}^{\rm(1-loop)}$ as given by Eq.\ \eqref{eq:tensor_inin_full}, i.e.\ the case of the diagram with two insertions of $H_{\gamma\zeta\zeta}$.
}
\label{fig:PBT}
\end{figure}

The result in \eqref{eq:tensor_inin_full} looks very small, suppressed both by a power of the slow-roll parameter and by the scalar power spectrum. But this is deceiving, since what we are really interested in is the relative correction to the tree-level result,
\begin{equation}
\mathcal{P}_\gamma^{\rm(tree)}=\frac{2}{\pi^2}\left(\frac{H}{\mpl}\right)^2=32\epsilon\cs(\rho\sin\psi)\mathcal{P}_\zeta e^{-2x} \,,
\end{equation}
so that the ratio $\mathcal{P}_\gamma^{\rm(1-loop)}/\mathcal{P}_\gamma^{\rm(tree)}$ is exponentially enhanced when $x$ is large. Fortunately, the ratio is still proportional to $\epsilon \mathcal{P}_\zeta$ and, at least on CMB scales, this can reasonably overcome the exponential term $e^{2x}$ as long as $x\lesssim\mathcal{O}(10)$ (here we are assuming as usual that $\rho\sin\psi$ is of order unity). We display the perturbativity bound, as function of $x$, $\cs$ and $\mathcal{P}_\zeta$, in Fig.\ \ref{fig:PBT}. 

%%%%%%%%%%%%%%%%%%%%%%%%%%%%%%%

\subsubsection*{Comment on higher loops.}

Given the dangerously large result for $\mathcal{P}_{\gamma}^{\rm(1-loop)}$ that we have found, it is interesting to naively ask if this correction might actually dominate over the tree-level result while still being consistent with the loop expansion. That is, we ask if \textit{higher} loop corrections could satisfy a consistent expansion even if $\mathcal{P}_\gamma^{\rm(1-loop)}/\mathcal{P}_\gamma^{\rm(tree)}>1$. A quick back-of-the-envelope estimate shows however that this is not the case, i.e.\ a violation of our perturbativity bound derived at one-loop order really signals a breakdown of the full loop expansion.

Consider for concreteness a two-loop diagram with four insertions of the cubic Hamiltonian $H_{\gamma\zeta\zeta}$. The two-point correlator is given schematically by
\begin{equation}
\braket{\gamma^2}^{\rm(2-loop)}\sim \mathscr E^4 \left\langle
\left(a^2\int \gamma\zeta\zeta \right)^2\gamma^2\left(a^2\int \gamma\zeta\zeta \right)^2
\right\rangle \sim \epsilon^4 \left(
\frac{\mpl}{H}
\right)^8\mathcal P_\zeta^4\mathcal P_\gamma^3 \,.
\end{equation}
Since $\braket{\gamma^2}^{\rm(1-loop)}\sim\epsilon^2 \mathcal P_\zeta^2\mathcal P_\gamma^2(\mpl/H)^4$, we have again that
\begin{equation}
\frac{\braket{\gamma^2}^{\rm(2-loop)}}{\braket{\gamma^2}^{\rm(1-loop)}}\sim \epsilon\mathcal P_\zeta e^{2x} \,.
\end{equation}
Similarly, for all higher loops, we can estimate $\mathcal{P}_\gamma^{((n+1)-\rm loop)}/\mathcal{P}_\gamma^{(n-\rm loop)}\sim\mathcal{P}_\gamma^{\rm(1-loop)}/\mathcal{P}_\gamma^{\rm(tree)}$, consistent with expectations. We caution that this quick estimate does not take into account factors of $n$, which could in principle lead to dangerous enhancements at large-loop order. See e.g.\ \cite{Firouzjahi:2025ihn} where such effect was observed.

%%%%%%%%%%%%%%%%%%%%%%%%%%%%%%%

\subsection{Loops with one quartic vertex}

It remains to consider the one-loop corrections from diagrams with one quartic vertex, given by the interaction Hamiltonians $H_{\gamma\gamma\zeta\zeta}^{(1)}$ and $H_{\gamma\gamma\zeta\zeta}^{(2)}$, cf.\ Eqs.\ \eqref{eq:quarticHam tensor1} and \eqref{eq:quarticHam tensor2}. It is easy to convince oneself, before doing any calculation, that these contributions are actually totally safe with regards to the large-$x$ expansion. Indeed, the in-in integral contains two fewer factors of $\zeta$ in comparison with the loop of the previous subsection, so we expect a suppression of $e^{-2x}$ in $\mathcal{P}_\gamma^{\rm(1-loop)}$ in the present case. Nevertheless, it turns out that loops with quartic vertices introduce a new behavior, namely an infrared divergence associated to the late-time limit.

%%%%%%%%%%%%%%%%%%%%%%%%%%%%%%%

\subsubsection*{$H_{\gamma\gamma\zeta\zeta}^{(1)}$-vertex.}

Inserting \eqref{eq:quarticHam tensor1} into the relevant in-in formula we obtain
\begin{equation} \label{eq:tensor_quartic1}
\braket{\gamma_{ij,\vec k}(\eta)\gamma_{ij,\vec p}(\eta)}=-\frac{8\mathscr F}{H^2}\delta^3(\vec k+\vec p)\int_{-\infty}^\eta\frac{\D\eta_1}{\eta_1^{2}}\int\D^3\vec p_1 |\zeta_{p_1}'(\eta_1)|^2\,\mathrm{Im}\left[\gamma_k(\eta_1)\gamma_k^*(\eta)\gamma_k(\eta_1)\gamma_k^*(\eta)\right] \,.
\end{equation}
Focusing as usual in the large-$x$ approximation, we take the scalar modes as growing, resulting in the following expression for the dimensionless one-loop tensor spectrum:
\begin{equation} \label{eq:tensor_quartic1a}
\frac{k^3}{2\pi^2}\braket{\gamma^2}'=
-\frac{8\mathcal F \epsilon \cs^2}{k^3}\mathcal P_\zeta\mathcal P_\gamma \,\mathrm{Im}\bigg[e^{2ik\eta}(i+k\eta)^2\int_{-\infty}^\eta \D\eta_1\int_0^{\infty} \D p_1\, p_1^3 e^{2p_1\cs\eta_1-2ik\eta_1}(i-k\eta_1)^2\bigg] \,.
\end{equation}
The upper limit of the momentum integral should in principle be cutoff at $p_1=-x/(\cs\eta_1)$. However, using a very similar argument as in the previous subsection, one can easily see that the integral in \eqref{eq:tensor_quartic1a} is actually convergent at $p_1\to\infty$. We conclude again that the leading-order result in the large-$x$ expansion is independent of $x$ and is precisely given by the above integral.

The integral \eqref{eq:tensor_quartic1a} is on the other hand IR divergent, i.e.\ it diverges as $\log(-k\eta)$ as $\eta\to0^-$,\footnote{We find it interesting that some non-trivial cancellations occur in order to produce this result. Indeed, the integral inside the imaginary part actually diverges as $\eta^{-3}$ and $\eta^{-1}$ as $\eta\to0^{-}$. However, all such dangerous powers turn out to be purely real and hence do not contribute to the final result.}
\begin{equation}
\frac{k^3}{2\pi^2}\braket{\gamma^2}'=\frac{2\epsilon\mathcal F\mathcal P_\zeta\mathcal P_\gamma}{\cs^2}\big[
    \log(-2k\eta)+\gamma_E-2
    \big]+\mathcal O(\eta^2) \,,
\end{equation}
where $\gamma_E$ is the Euler constant. Such secular divergences, arising from non-linear effects accumulating on super-Hubble scales, are commonplace in perturbative QFT in de Sitter space.
However, there is never any infrared divergent correlator in actual inflationary models, simply because inflation had a finite duration; or equivalently, these divergences may be seen as artifacts of assuming perfect scale invariance.
Hence it is sufficient for our purpose to simply regularize the result by introducing an IR cutoff $-k\eta_{\rm IR}=\Lambda_{\rm IR}/H$, corresponding to a time $\eta_{\rm IR}$ long after the mode $k$ of interest exits the horizon, with the assumption $\Lambda_{\rm IR}\ll H$. Evaluating the result one arrives at
\begin{equation}
\mathcal{P}_\gamma^{\rm(1-loop)}\simeq -\frac{2\epsilon \mathcal F\mathcal P_\zeta\mathcal P_\gamma}{\cs^2}\log \frac{H}{\Lambda_{\rm IR}} \,,
\end{equation}
where the approximate sign is because we are neglecting subleading terms in the large-$x$ expansion, which are at least proportional to $1/x$, as well as IR-finite terms, i.e.\ terms which converge as $\Lambda_{\rm IR}\to 0$.

As claimed, the ratio $\mathcal{P}_\gamma^{\rm(1-loop)}/\mathcal{P}_\gamma^{\rm(tree)}$ is independent of $x$ at leading order, so this one-loop correction can be considered as safe from the point of view of perturbativity: even for reasonably large values of $H/\Lambda_{\rm IR}$, the suppression by $\epsilon\mathcal P_\zeta\ll1$ should ensure perturbative control.

%%%%%%%%%%%%%%%%%%%%%%%%%%%%%%%

\subsubsection*{$H_{\gamma\gamma\zeta\zeta}^{(2)}$- and $H_{\gamma\gamma\zeta\zeta}^{(3)}$-vertices.}

The loop corrections from the second and third interactions, Eqs.\ \eqref{eq:quarticHam tensor1} and \eqref{eq:quarticHam tensor3}, can be computed following the exact same procedure, with analogous results: finite in the large-$x$ limit but logarithmically IR-divergent. We quote here only the final results:
\begin{equation}
\begin{aligned}
\mathcal{P}_\gamma^{\rm(1-loop)}&\simeq -\frac{6\epsilon\tilde{\mathcal F}\mathcal P_\zeta\mathcal P_\gamma}{\cs^2} \log\frac{H}{\Lambda_{\rm IR}} \,,\\
\mathcal{P}_{\gamma}^{\rm (1-loop)}&\simeq -\frac{\epsilon \overline{\mathcal{F}}\mathcal{P}_\zeta \mathcal{P}_{\gamma}}{\cs^2}\log \frac{H}{\Lambda_{\rm IR}} \,,
\end{aligned}
\end{equation}
respectively for the $H_{\gamma\gamma\zeta\zeta}^{(2)}$ and $H_{\gamma\gamma\zeta\zeta}^{(3)}$ vertices.

%%%%%%%%%%%%%%%%%%%%%%%%%%%%%%%%%%
%%%%%%%%%%%%%%%%%%%%%%%%%%%%%%%%%%

\section{Discussion}
\label{sec:discussion}

Our main aim in this paper was to address the question of whether the EFT of inflation with imaginary speed of sound admits a consistent loop expansion within the space of parameters of phenomenological interest, i.e.\ those found in concrete multi-field realizations such as strongly non-geodesic models of inflation characterized by a large and negative entropic mass $m_s^2$.

At least generically, one expects the ratio of the one-loop correction of an observable to its tree-level value to give a good estimate of the parameter controlling the full loop expansion, cf.\ our discussion in Sec.\ \ref{subsec:tensor cubic}. The simplest such observables in the present context are the scalar and tensor power spectra, which we treated in this paper, respectively in Sec.\ \ref{sec:one loop scalar spectrum} and Sec.\ \ref{sec:one loop tensor spectrum}. Interestingly, we have found qualitatively distinct results depending on the type of one-loop diagram:
\begin{itemize}
    \item Loop corrections to the scalar spectrum from diagrams with two cubic vertices provide the dominant contribution, of order $\mathcal{P}_{\zeta}^{\rm(1-loop)}\sim \mathcal{P}_{\zeta}^2x^5$, ignoring numerical factors and $\cs\sim1$. For small $\cs$, this is further enhanced by $\cs^{-4}$;
    \item Loop corrections to the scalar spectrum from diagrams with one quartic vertex give a subleading contribution, of order $\mathcal{P}_{\zeta}^{\rm(1-loop)}\sim \mathcal{P}_{\zeta}^2x$ if $\cs\sim1$, enhanced by $\cs^{-4}$ if $\cs$ is small;
    \item Loop corrections to the tensor spectrum from diagrams with two cubic vertices yield a dangerous, exponentially enhanced correction of order $\mathcal{P}_{\gamma}^{\rm(1-loop)}/\mathcal{P}_{\gamma}^{\rm(tree)}\sim \epsilon\mathcal{P}_{\zeta}e^{2x}$;
    \item Loop corrections to the tensor spectrum from diagrams with one quartic vertex are not at all enhanced by powers or exponentials of $x$. However, they exhibit a qualitative difference: they are logarithmically IR-divergent at late times.
\end{itemize}
As a technical aspect, our calculations required a careful manipulation of loop integrals, which in the EFT with imaginary sound speed must be regularized with a (time-dependent) cutoff on momenta. This procedure results in a non-trivial domain of integration for nested time and momentum integrals, implying in particular that care is needed when exchanging the order of integration. We have striven to describe this prescription in detail, as we expect that it may find application in other contexts involving loop calculations within the in-in formalism.

As mentioned repeatedly throughout the paper, our results and conclusions rely on the large-$x$ approximation, which is a necessary condition for the EFT to capture the sub-sound horizon physics that produces the salient features of this class of models. In practice, it has been shown that $x=\mathcal{O}(10)$ is typically large enough for the EFT to correctly reproduce the predictions of the multi-field theory, at least in cases where the comparison can be investigated in detail~\cite{Fumagalli:2019noh}. Moreover, and importantly, $x$ cannot be arbitrarily large, since in fact one has the bound $x\lesssim |m_s|/H$ as a requirement for the consistency of integrating out a heavy tachyonic entropic field~\cite{Garcia-Saenz:2018vqf}. For instance, in the models of sidetracked inflation~\cite{Garcia-Saenz:2018ifx} with large entropic mass, one has $\cs\ll1$ and $x\sim \eta_\perp\sim|m_s|\cs/H$ on dimensional grounds (see also \cite{Bjorkmo:2019qno}), consistent with the above bound.

It is worth commenting on the accuracy of our results. The first remark is that our regularization prescription, in which momentum integrals are bounded by a hard cutoff, is expected to provide a correct order of magnitude, although not necessarily an accurate result whenever there is a dependence on the cutoff. Remarkably, in this regard, our main result concerning the dominant contribution to the tensor spectrum is that this observable is actually independent of the cutoff, which we therefore expect to be more accurate than our corresponding results for the scalar power spectrum. 
The second comment concerns the possibility that some of our results may be potentially canceled by counterterms when the theory is renormalized. We have given several arguments in support of the tenet that our results do capture physical effects. Nevertheless, a more thorough answer to this question would require a rigorous understanding of how renormalization works in this theory, which we currently lack. Also interesting, although technically challenging, would be to compare our results with the calculation performed in a multi-field UV completion, following for instance the techniques developed in \cite{Fumagalli:2021mpc}.

An important conclusion is that loop corrections can easily become dangerously large in the EFT with imaginary speed of sound. This is relevant both in the formal context of the EFT itself, but also in the context of strongly non-geodesic models of inflation, given that the question of perturbativity in the multi-field theory may be reliably diagnosed within the EFT, assuming of course one works within its regime of validity. The issue of large loops concerns particularly scenarios predicting a very large bending, say $x\gtrsim 100$, or ones that feature a strong enhancement of the scalar power spectrum on small scales. For instance, taking into consideration the perturbativity of the scalar power spectrum, cf.\ Fig.\ \ref{fig:PB}, we find that very large values of $\mathcal{P}_\zeta$, e.g.\ $\mathcal{P}_\zeta\sim 10^{-2}$, appear to be inconsistent as soon as $x\gtrsim 10$, barring cancellations among different diagrams, a situation that is only worsened for small values of $\cs$.\footnote{This does not immediately rule out large values of $\mathcal{P}_\zeta$ if they were to also exhibit a strong scale dependence. It is worth recalling that all our results rest on the assumption of approximate scale-invariance.} On the other hand, it is an interesting result that this tension is mostly absent in models with $\cs=\mathcal{O}(1)$, at least provided $x$ is not too large, which as we have remarked is a reasonable situation and is found in concrete multi-field realizations. It would be interesting to consider loop corrections to other observables to see whether our bounds can be tightened even further.

\subsubsection*{Acknowledgments}

SGS acknowledges support from the NSFC (Grant No.\ 12250410250) and from a provincial grant (Grant No.\ 2023QN10X389). The work of YL was supported by the NSFC (Grant No.\ 12247161), by the China Postdoctoral Science Foundation (Grant No.\ 2022TQ0140) and by the SIMIS. During the course of this work, S.RP was supported by the European Research Council under the European Union's Horizon 2020 research and innovation programme (grant agreement No.\ 758792, Starting Grant project GEODESI).

\appendix

\section{Lowest-order EFT and k-inflation}\label{app:k-inf}

We mentioned that our set-up is equivalent to the theory of fluctuations in the context of k-inflation \cite{Garriga:1999vw,Armendariz-Picon:2000nqq}, i.e.\ the covariant scalar field model defined by a Lagrangian of the form $P(X,\phi)$, with $X\equiv -\frac{1}{2}(\nabla\phi)^2$, and $P$ an arbitrary function. Although well known, this equivalence is not manifest due to the usual ambiguities related to boundary terms and field redefinitions, so we provide a brief review of this aspect here.

The cubic action for $\zeta$ in the EFT at lowest order in derivatives and in the slow-roll expansion is given by \cite{Cheung:2007st,Cheung:2007sv}
\begin{equation}\label{eq:cubic_kinf}
    S_{\zeta^3}=\int\D t\D^3 \vec x \,a^3\frac{\epsilon\mpl^2}{H}\left(1-\frac1{c_s^2}\right)\left[
    \frac{\mathcal A}{c_s^2}\dot\zeta^3+\dot\zeta\frac{(\partial\zeta)^2}{a^2}
    \right] \,.
\end{equation}
We warn the reader that we use cosmic time $t$ in this Appendix, unlike in the rest of the paper. Actually, we recall that the EFT of inflation starts life not with the curvature perturbation $\zeta$, but with the Nambu-Goldstone boson of broken time translations, denoted by $\pi$. The two are related non-linearly as $\zeta=-H\pi+H\pi\dot{\pi}+\frac{1}{2}\dot{H}\pi^2+\mathcal{O}(\pi^3)$ \cite{Maldacena:2002vr}. This point is immaterial as far as the cubic Lagrangian is concerned (substituting the non-linear terms in the quadratic action for $\zeta$ will generate cubic terms, but these are proportional to the linear equation of motion, which are irrelevant in a perturbative treatment), but it is important if one wishes to determine the precise coefficients of the quartic vertices. As explained in the main text, in this paper we were not concerned with calculating these coefficients (e.g.\ in terms of the defining coefficients of the EFT) but rather considered them as extra free parameters.

The standard form of the cubic action for $\zeta$ derived in k-inflation is given by \cite{Chen:2006nt}
\begin{equation} \label{eq:chencubic}
S_{\zeta^3}=\int \D\eta\D^3\vec x\,a^3
\frac{\epsilon M_{\rm Pl}^2}{c_s^2}
\bigg\{-\left[\left(1-\frac1{c_s^2}\right)+\frac{2\lambda}\Sigma\right]\frac{\dot\zeta^ 3}{H}+3\left(1-\frac1{c_s^2}\right)\zeta\dot\zeta^{2}+\left(1-c_s^2\right)\zeta\frac{(\partial\zeta)^2}{a^2}
\bigg\} \,.
\end{equation}
remembering that we neglect subleading corrections in the slow-roll expansion. Here $\Sigma\equiv XP_X+2X^2P_{XX}=\epsilon H^2/c_s^2$ and $\lambda\equiv X^2P_{XX}+\frac{2}{3}X^3P_{XXX}$ are defined in terms of $X$ and partial derivatives of the function $P$ (e.g.\ $P_X\equiv \frac{\partial P}{\partial X}$), all evaluated on the background. Although not obvious, \eqref{eq:cubic_kinf} and \eqref{eq:chencubic} coincide up to boundary terms, with the identification $\mathcal A(1-c_s^{-2})=2\lambda/\Sigma$ \cite{Renaux-Petel:2011zgy,Garcia-Saenz:2019njm}.\footnote{Although here we are focusing only on the leading terms in the slow-roll expansion, we emphasize that the equivalence is valid at all orders in slow-roll. Incidentally, the relation $\mathcal A(1-c_s^{-2})=2\lambda/\Sigma$ is exact and does not assume any slow-roll approximation.}

The difference between \eqref{eq:cubic_kinf} and \eqref{eq:chencubic} is given explicitly by
\begin{equation}\label{eq:app-boundary}
    -\int\D t\D^3\vec x \,\frac{\D}{\D t}\bigg[
    \frac{a\epsilon\mpl^2}{Hc_s^2}(1-c_s^2)\zeta(\partial\zeta)^2+\frac{a^3\epsilon\mpl^2}{Hc_s^4}(1-c_s^2)\zeta\dot\zeta^2
    \bigg]+\ldots \,,
\end{equation}
where the ellipses stand for \textit{spatial} boundary terms, which do not affect cosmological correlators. The first term inside the brackets does not contain $\dot{\zeta}$ and therefore does not contribute to correlation functions that include only fields (and not field momenta) \cite{Burrage:2011hd,Braglia:2024zsl,Fumagalli:2024jzz}, while the second term does not contribute to these correlators because of the super-Hubble conservation of $\zeta$.

%%%%%%%%%%%%%%%%%%%%%%%%%%%%%%%%%%%
%%%%%%%%%%%%%%%%%%%%%%%%%%%%%%%%%%%

\section{Scalar loop from $H_{\zeta\zeta\zeta}^{(2)}$}
\label{app:H2 scalar loop}

The calculation of the one-loop scalar spectrum with two insertions of the $H_{\zeta\zeta\zeta}^{(2)}$ vertex (cf.\ Eq.\ \eqref{eq:cubicHam2}) is made lengthy by the number of inequivalent permutations. In this Appendix we provide details of the computation which we preferred to avoid in the main text.

\begin{figure}
\centering
\includegraphics[width=\textwidth]{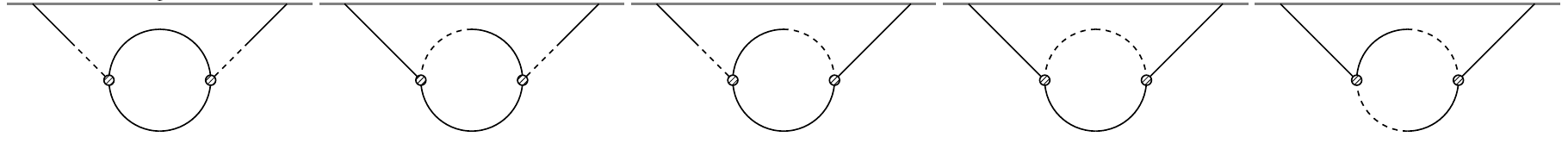}
\caption{Five permutations for the one-loop diagram with two insertions of the cubic vertex $H_{\zeta\zeta\zeta}^{(2)}$.
Dashed curves represent modes with time derivative.}
\label{fig:cubic2loop}
\end{figure}

There are five distinct permutations that contribute to the loop integral, as shown in Fig.\ \ref{fig:cubic2loop}; we label these, from left to right, as $a,b,c,d,e$. The individual contributions to the two-point correlator are given by
\begin{equation}
\begin{aligned}
\braket{\zeta_{\vec k}(\eta)\zeta_{\vec p}(\eta)}^{(a)}&=-\frac{16 \tilde{\mathscr C}^2}{H^{2}}\delta^3(\vec k+\vec p)\int \mathrm{d}\eta_1 \eta_1^{-1}\int\mathrm{d}\eta_2 \eta_2^{-1}\int \mathrm{d}^3\vec p_1 \mathrm{d}^3\vec p_2 \,\delta^3(\vec p_1+\vec p_2-\vec k) \\
&\quad\times(\vec p_1\cdot \vec p_2)^2
\mathrm{Im}\left[\zeta_{k}'(\eta_1)\zeta_{k}^*(\eta)\right]\mathrm{Im}\left[\zeta_{k}(\eta)\zeta_{k}^{*\prime}(\eta_2)\zeta_{p_1}(\eta_1)\zeta_{p_1}^{*}(\eta_2)\zeta_{p_2}(\eta_1)\zeta_{p_2}^{*}(\eta_2)\right] \\
&=\frac{2\pi^2}{k^3}\delta^3(\vec k+\vec p)\frac{8\tilde{\mathscr C}^2\cs^2}{\pi H^2}A^6\mathrm{Im}[B]^2e^{2k\cs\eta}(k\cs\eta-1)^2 \\
&\quad\times\left[
\frac{x^5}{150}(2\log 2-1)+\frac{x^4}{12}(2\log2-1)+\frac{x^3}{12}(8\log 2-5)+\mathcal O(x^2)
\right] \,,
\end{aligned}
\end{equation}
\begin{equation}
\begin{aligned}
\braket{\zeta_{\vec k}(\eta)\zeta_{\vec p}(\eta)}^{(b)}&=-\frac{32 \tilde{\mathscr C}^2}{H^{2}}\delta^3(\vec k+\vec p)\int \mathrm{d}\eta_1 \eta_1^{-1}\int  \mathrm{d}\eta_2 \eta_2^{-1}\int \mathrm{d}^3 \vec p_1 \mathrm{d}^3 \vec p_2\,\delta^3(\vec p_1+\vec p_2-\vec k) \\
&\!\!\!\!\times(\vec p_1\cdot \vec p_2)(\vec k\cdot \vec p_2)
\mathrm{Im}\left[\zeta_{k}(\eta_1)\zeta_{k}^*(\eta)\right]
\mathrm{Im}\left[\zeta_{k}(\eta)\zeta_{k}^{*\prime}(\eta_2)\zeta_{p_1}'(\eta_1)\zeta_{p_1}^{*}(\eta_2)\zeta_{p_2}(\eta_1)\zeta_{p_2}^{*}(\eta_2)\right] \\
&=\frac{2\pi^2}{k^3}\delta^3(\vec k+\vec p)\frac{16\tilde{\mathscr C}^2\cs^2}{\pi H^2}A^6\mathrm{Im}[B]^2e^{2k\cs\eta}(k\cs\eta-1)^2\\
&\quad\times\left[
\frac{x^5}{150}(2\log 2-1)+\frac{13x^4}{240}(2\log2-1)+\frac{x^3}{720}(224\log 2-197)+\mathcal O(x^2)
\right] \,,
\end{aligned}
\end{equation}
\begin{equation}
\begin{aligned}
\braket{\zeta_{\vec k}(\eta)\zeta_{\vec p}(\eta)}^{(c)}&=-\frac{32 \tilde{\mathscr C}^2}{H^{2}}\delta^3(\vec k+\vec p)\int \mathrm{d}\eta_1 \eta_1^{-1}\int  \mathrm{d}\eta_2 \eta_2^{-1}\int \mathrm{d}^3 \vec p_1 \mathrm{d}^3 \vec p_2\,\delta^3(\vec p_1+\vec p_2-\vec k) \\
&\!\!\!\!\times(\vec p_1\cdot \vec p_2)(\vec k\cdot \vec p_2)
\mathrm{Im}\left[\zeta_{k}'(\eta_1)\zeta_{k}^*(\eta)\right]
\mathrm{Im}\left[\zeta_{k}(\eta)\zeta_{k}^{*}(\eta_2)\zeta_{p_1}(\eta_1)\zeta_{p_1}^{*\prime}(\eta_2)\zeta_{p_2}(\eta_1)\zeta_{p_2}^{*}(\eta_2)\right] \\
&=\frac{2\pi^2}{k^3}\delta^3(\vec k+\vec p)\frac{16\tilde{\mathscr C}^2\cs^2}{\pi H^2}A^6\mathrm{Im}[B]^2e^{2k\cs\eta}(k\cs\eta-1)^2\\
&\quad\times\left[
\frac{x^5}{150}(2\log 2-1)-\frac{x^4}{120}(\log2-4)+\frac{x^3}{720}(151-208\log 2)+\mathcal O(x^2)
\right] \,,
\end{aligned}
\end{equation}
\begin{equation}
\begin{aligned}
\braket{\zeta_{\vec k}(\eta)\zeta_{\vec p}(\eta)}^{(d)}&=-\frac{32 \tilde{\mathscr C}^2}{H^{2}}\delta^3(\vec k+\vec p)\int \mathrm{d}\eta_1 \eta_1^{-1}\int  \mathrm{d}\eta_2 \eta_2^{-1}\int \mathrm{d}^3 \vec p_1 \mathrm{d}^3 \vec p_2\,\delta^3(\vec p_1+\vec p_2-\vec k) \\
&\quad\times(\vec k\cdot \vec p_2)^2
\mathrm{Im}\left[\zeta_{k}(\eta_1)\zeta_{k}^*(\eta)\right]
\mathrm{Im}\left[\zeta_{k}(\eta)\zeta_{k}^{*}(\eta_2)\zeta_{p_1}'(\eta_1)\zeta_{p_1}^{*\prime}(\eta_2)\zeta_{p_2}(\eta_1)\zeta_{p_2}^{*}(\eta_2)\right] \\
&=\frac{2\pi^2}{k^3}\delta^3(\vec k+\vec p)\frac{16\tilde{\mathscr C}^2\cs^2}{\pi H^2}A^6\mathrm{Im}[B]^2e^{2k\cs\eta}(k\cs\eta-1)^2\\
&\quad\times\left[
\frac{x^5}{150}(2\log 2-1)-\frac{x^4}{240}(16\log2-15)+\frac{x^3}{120}(8\log 2-11)+\mathcal O(x^2)
\right] \,,
\end{aligned}
\end{equation}
\begin{equation}
\begin{aligned}
\braket{\zeta_{\vec k}(\eta)\zeta_{\vec p}(\eta)}^{(e)}&=-\frac{32 \tilde{\mathscr C}^2}{H^{2}}\delta^3(\vec k+\vec p)\int \mathrm{d}\eta_1 \eta_1^{-1}\int  \mathrm{d}\eta_2 \eta_2^{-1}\int \mathrm{d}^3 \vec p_1 \mathrm{d}^3 \vec p_2\,\delta^3(\vec p_1+\vec p_2-\vec k) \\
&\!\!\!\!\times(\vec k\cdot \vec p_1)(\vec k\cdot \vec p_2)
\mathrm{Im}\left[\zeta_{k}(\eta_1)\zeta_{k}^*(\eta)\right]
\mathrm{Im}\left[\zeta_{k}(\eta)\zeta_{k}^{*}(\eta_2)\zeta_{p_1}'(\eta_1)\zeta_{p_1}^{*}(\eta_2)\zeta_{p_2}(\eta_1)\zeta_{p_2}^{*\prime}(\eta_2)\right] \\
&=\frac{2\pi^2}{k^3}\delta^3(\vec k+\vec p)\frac{16\tilde{\mathscr C}^2\cs^2}{\pi H^2}A^6\mathrm{Im}[B]^2e^{2k\cs\eta}(k\cs\eta-1)^2\\
&\quad\times\left[
\frac{x^5}{150}(2\log 2-1)-\frac{x^4}{240}(16\log2-15)-\frac{x^3}{360}(56\log 2-17)+\mathcal O(x^2)
\right] \,.
\end{aligned}
\end{equation}
Adding all these contributions yields the result quoted in the main text, Eq.\ \eqref{eq:1loop_cubic2}.

%%%%%%%%%%%%%%%%%%%%%%%%%%%%%%%%%%%
%%%%%%%%%%%%%%%%%%%%%%%%%%%%%%%%%%%

\section{Other quartic scalar vertices}
\label{app:other quartic}

In the main text we considered, for simplicity, only a single quartic $\zeta$ vertex, i.e.\ $\zeta^{\prime4}$. The resulting one-loop correction to the scalar power spectrum was found to scale as $x$ in the limit of large $x$, and is thus sub-dominant in comparison to the contributions from cubic vertices. In this Appendix we argue that the same scaling should appear in the results for the other two scalar vertices, i.e.\ $\zeta^{\prime2}(\partial\zeta)^2$ and $(\partial\zeta)^4$. We will carry out explicit calculations for the latter vertex, although it will become apparent that the same argument may be used for the other one in order to arrive at the same conclusion.

\begin{figure}
    \centering
    \includegraphics{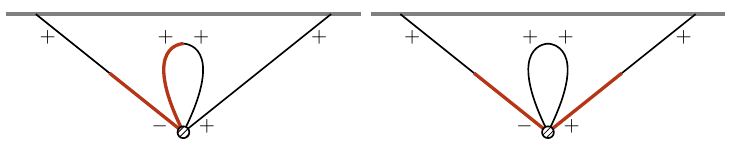}
    \caption{Two permutations of the one-loop diagram for the quartic vertex \eqref{eq:app_quartic2}.
    Red lines indicate one pair $(\partial\zeta)^2$ out of the two in the vertex.}
    \label{fig:quartic2}
\end{figure}

Consider then for concreteness the interaction Hamiltonian
\begin{equation}\label{eq:app_quartic2}
H_{\zeta\zeta\zeta\zeta}^{(2)}(\eta)=\tilde{\mathscr D}\int\mathrm d^3\vec x\, (\partial\zeta)^4 \,,\qquad \tilde{\mathscr D}\equiv\frac{M_{\rm Pl}^2\epsilon}{H^2|c_s|^2}\mathcal{B} \,,
\end{equation}
where $\mathcal{B}$ is a dimensionless constant, expected to be $\mathcal{O}(1)$ in the EFT context. At leading order in $x$, i.e.\ upon maximizing the number of growing modes, in the loop integral, one has the diagram shown in Fig.\ \ref{fig:quartic2}, where we have separated the two types of permutations: one where both factors of $(\partial\zeta)^2$ share an external leg and internal one, and one where one factor of $(\partial\zeta)^2$ is purely external and the other purely internal. The former permutation contributes a factor of $(\vec k\cdot \vec p)^2$, where $\vec k$ is the external momentum and $\vec p$ is the internal one, while the latter contributes as $k^2p^2$. Since $(\vec k\cdot \vec p)^2\leq k^2p^2$, let us focus first only on the latter permutation for the purpose of estimating the size of the loop integral:
\begin{equation} \label{eq:app_B_spatial}
\tilde{\mathscr {I}}_2=\int_{-x/k\cs}^\eta\D\eta_1\int_0^{-x/\cs\eta_1}\D p\,
{\color{LouisBlue}k^2p^2}\tilde{g}(p,\eta_1)e^{2p\cs\eta_1},
\end{equation}
where the {\color{LouisBlue}blue} factor is specific to the form of the $(\partial\zeta)^4$ vertex, as we just explained, and the function $\tilde{g}$ encapsulates all other contributions (mode functions, coupling constants, etc.). We wish to compare \eqref{eq:app_B_spatial} with the equivalent expression for the $\zeta^{\prime4}$ vertex, see Eq.\ \eqref{eq:I1 integral}. Noting that
\begin{equation}
\zeta'_{p,\pm}=\pm \frac{p^2\cs^2\eta}{p\cs\eta-1}\zeta_{p,\pm} \,,
\end{equation}
it is straightforward to see that the dominant $\zeta^{\prime4}$ loop integral is
\begin{equation} \label{eq:app_B_time}
\mathscr I_2=\int_{-x/k\cs}^\eta\D\eta_1\int_0^{-x/\cs\eta_1}\D p\,
{\color{LouisBlue}\frac{-p^4\cs^8\eta_1^4}{(p\cs\eta_1-1)^2(k^2\cs^2\eta_1^2-1)}}g(p,\eta_1)e^{2p\cs\eta_1} \,.
\end{equation}
The functions $\tilde g(p,\eta_1)$ and $g(p,\eta_1)$ are the same up to coupling constants and order-one factors, so the point here is to compare the {\color{LouisBlue}blue} terms in Eqs.\ \eqref{eq:app_B_spatial} and \eqref{eq:app_B_time}. One notes that both integrals are dominated by the region where $-p\cs\eta_1\sim \mathcal O(1)$, and in fact the leading $x$ behavior is obtained by setting $p\sim -1/\cs\eta_1$ and $\eta_1\sim -x/k\cs$, for which both {\color{LouisBlue}blue} factors give the same scaling, namely $\propto x^{-2}$, and so both integrals will ultimately also scale in the same way, namely $\propto x$. The other vertex, $\zeta^{\prime2}(\partial\zeta)^2$, may be checked to produce the same result following an analogous argument.

As a final verification, we have performed the full calculation of the one-loop correction to the power spectrum from the $(\partial\zeta)^4$ vertex, with the result
\begin{equation}
\frac{k^3}{2\pi^2}\langle\zeta^2\rangle^{\prime}_{(\partial\zeta)^4}=\mathcal{P}_\zeta^2\frac{\mathcal{B}}{\cs^4}e^{2k\cs\eta}(k\cs\eta-1)^2 \left(9x+\mathcal O(x^0)\right) \,,
\end{equation}
confirming the $\propto x$ scaling of our quick estimate. 

%%%%%%%%%%%%%%%%%%%%%%%%%%%%%%%
%%%%%%%%%%%%%%%%%%%%%%%%%%%%%%%

%==========================

\bibliographystyle{JHEP}
\bibliography{ref}
\end{document}